\documentclass[onecolumn,pra,nofootinbib]{revtex4}
\usepackage{amsmath}      %
\usepackage{amssymb}      
\usepackage{amsfonts}     %
\usepackage{graphicx}     
\usepackage{hyperref}
\begin{document}
\title{Non-linear uplift Ans\"atze for the internal metric and the four-form
  field-strength of maximal supergravity}
\author{Olaf Kr\"uger}
\affiliation{Max-Planck-Institut f\"ur Gravitationsphysik,
  Albert-Einstein-Institut, Am M\"uhlenberg 1, D-14476 Potsdam,
  Germany}
\date{\today}
\begin{abstract}
  The uplift of SO(8) gauged $N=8$ supergravity to 11-dimensional
  supergravity is well studied in the literature. It is given by
  consistent relations between the respective vector and scalar fields
  of both theories. For example, recent work provided non-linear
  uplift Ans\"atze for the scalar degrees of freedom on the internal
  manifold: the inverse metric and the three-form flux with mixed
  index structure. However, one always found the metric of the
  compactified manifold by inverting the inverse metric --- a task
  that was only possible in particular cases, e.g. for the G$_2$,
  SO(3)$\times$SO(3) or SU(3)$\times$U(1)$\times$U(1) invariant
  solutions of 11-dimensional supergravity.

  In this paper, I present a direct non-linear uplift Ansatz for the
  internal metric in terms of the four-dimensional scalars and the
  Killing forms on the compactified background manifold. Based on this
  formula, I also find new uplift Ans\"atze for the warp factor and
  the full internal three-form flux, as well as for the internal
  four-form field-strength. The new formula for the four-form only
  depends on the metric, the flux as well as the four-dimensional
  scalars and background Killing forms --- it does not require to
  calculate the derivative of the flux. All the Ans\"atze presented in
  this work pass a very non-trivial test for a G$_2$ invariant
  solution of 11-dimensional supergravity.

  My results may be generalized to other compactifications, e.g. the
  reduction from type IIB supergravity to five dimensions.
\end{abstract}
\maketitle

\section{Introduction}
\label{sec:introduction}

A supergravity theory in $D>4$ dimensions may be related to a
four-dimensional theory of gravity coupled to matter. This is the idea
of \textit{Kaluza-Klein theory}: A $D$-dimensional manifold splits
into a four-dimensional and a compact $(D-4)$-dimensional manifold,
\begin{equation}
  \label{eq:compactification}
  \mathcal{M}_D = \mathcal{M}_4 \times \mathcal{M}_{D-4}.
\end{equation}
This splitting is called \textit{compactification} of the $(D-4)$
extra dimensions. An action including the $D$-dimensional
Einstein-Hilbert term is given by
\begin{equation}
  \label{eq:11d-action}
  S = \int (R_D + \ldots) \, \mathrm{d}V,
\end{equation}
where $R_D$ denotes the Ricci scalar in $D$ dimensions. For a
consistent compactification, Eq.~(\ref{eq:11d-action}) contains the
four-dimensional Einstein-Hilbert action. All other terms correspond
to matter. For example, T. Kaluza and O. Klein presented one of the
first attempts to unify gravity and electromagnetism
\cite{Kaluza:1921tu,Klein:1926}. They constructed a five-dimensional
theory of gravity,
\begin{equation}
  \label{eq:Kaluza-Klein-Action}
  S_5 = \int R_5 \mathrm{d}V
\end{equation}
such that the extra components of the metric were given by a photon
and a scalar field. In that case, the fifth dimension was compactified
on a circle,
\begin{equation}
  \label{eq:Kaluza-Klein-Compactification}
  \mathcal{M}_5 = \mathcal{M}_4 \times S^1.  
\end{equation}

A physicist naturally is in another situation. He `observes' a
four-dimensional theory of gravity coupled to matter and may ask the
following question: Is there a higher-dimensional theory, which
consistently reduces to the observed theory via compactification of
the extra dimensions? This is called an \textit{uplift}: One
constructs the $D$-dimensional fields (e.g. the metric) out of a given
four-dimensional theory of gravity. The main task in establishing such
a program is to find Ans\"atze for the $D$-dimensional fields in terms
of the four-dimensional ones, such that they satisfy the
higher-dimensional equations of motion. The uplift is
\textit{consistent} only when the latter is satisfied.

One of the few known examples is the uplift of $N=8$ supergravity to
11-dimensional supergravity. $N=8$ supergravity represents the
low-energy limit of string theory. It is the maximally supersymmetric
theory of gravity and contains a local SU(8) gauge symmetry. It was
first investigated in the beginning of the 80s
\cite{cremmerjulia,deWit:1982ig}. At the same time, 11-dimensional
supergravity was developed \cite{Cremmer:1978km}, which is the highest
dimensional supergravity theory \cite{Nahm:1977tg}. The respective
Lagrangian is also locally SU(8) gauge invariant.

11-dimensional supergravity may spontaneously compactify to SO(8)
gauged $N=8$ supergravity
\cite{Englert:1982vs,Biran:1982eg,deWit:1986iy,Duff:1983gq}. The seven
extra dimensions therefore compactify on a seven-sphere\footnote{SO(8)
  is the isometry group for $S^7$.},
\begin{equation}
  \label{eq:11d-compactification}
  \mathcal{M}_{11} = \mathcal{M}_4 \times S^7.
\end{equation}

This work is based on the \textit{uplift} of SO(8) gauged $N=8$
supergravity to 11-dimensional supergravity
\cite{deWit:1986iy,NP,dWN13,Godazgar:2013dma,Godazgar:2013pfa}. It is
given by non-linear Ans\"atze for the 11-dimensional scalar and vector
fields in terms of the four-dimensional ones. These include the
correct relations between the 28 vector fields of 11-dimensional
supergravity and the 28 vectors of $N=8$ supergravity. On the other
hand, the 70 scalar degrees of freedom of 11-dimensional supergravity
are contained in certain fields that are defined on the internal space
(a \textit{deformed} seven-sphere): the metric $g_{mn}$, the
three-form potential $A_{mnp}$ and the six-form potential
$A_{m_1\cdots m_6}$. For the complete uplift, these fields must be
related to the 35 scalars $u_{ij}{}^{IJ}$ and pseudo-scalars
$v_{ij\,IJ}$ of $N=8$ supergravity.

There is an old explicit formula for the inverse metric
$\Delta^{-1}g^{mn}$ \cite{deWit:1984nz}, as well as non-linear
Ans\"atze for the full internal six-form potential and the three-form
flux with mixed index-structure \cite{Godazgar:2013pfa}. There are two
technical problems arising here: First, one must invert $\Delta^{-1}
g^{mn}$ `by hand' in order to obtain $\Delta g_{mn}$. Secondly, one
must extract the warp factor $\Delta$ from these expressions by
computing their determinants. Both, the inversion of the metric and
the calculation of the warp factor can only be done in particular
cases, e.g. when the theory is G$_2$, SO(3)$\times$SO(3) or
SU(3)$\times$U(1)$\times$U(1) invariant
\cite{Godazgar:2013nma,Godazgar:2014eza,Pilch:2015vha,Pilch:2015dwa}.
Only in such cases, it is then possible to compute the full internal
three-form potential $A_{mnp}$.

\textbf{In this paper, I present a new simple non-linear Ansatz for
  the full internal metric $g_{mn}$, i.e.}
\begin{equation}
  \label{eq:metric-Intro}
  \Delta^{-2} g_{mn}(x,y) = \,\frac{1}{12}
  \left( \mathcal{A}_m{}_{ijkl} -
    \mathcal{B}_m{}_{ijkl} 
  \right)
  \left(
    \mathcal{A}_n{}^{ijkl} -
    \mathcal{B}_n{}^{ijkl}
  \right)(x,y).
\end{equation}
The tensors $\mathcal{A}_m{}^{ijkl}$ and $\mathcal{B}_m{}^{ijkl}$ are
given in terms of the Killing forms on the seven-sphere and the
four-dimensional scalar fields
\textbf{(}Eqs.~(\ref{eq:pre-V}-\ref{eq:B})\textbf{)}. In combination
with the previous uplift formulas for the inverse metric and the
three-form with mixed index structure, \textbf{I also find new
  non-linear Ans\"atze for the warp factor and the full internal
  three-form potential $A_{mnp}$.} They are given by
\begin{gather}
  \label{eq:warp-Intro}
  \Delta^{-3}(x,y) = \frac{1}{28 \cdot 4!}
  \,\mathcal{C}_{ij}{}^{klmn}(x,y) \mathcal{C}^{ij}{}_{klmn}(x,y),\\
  \label{eq:A3-Intro}
  \Delta^{-3} A_{mnp}(x,y) = - \frac{\sqrt{2}i}{48\cdot 4!} K_{mn}{}^{IJ}(y)
  \left( u^{ij}{}_{IJ} -
    v^{ij\;IJ}
  \right)(x) \, \mathcal{C}_{ij}{}^{qrst} (x,y) \left( \mathcal{A}_{p\,qrst} -
    \mathcal{B}_{p\,qrst}\right) (x,y),
\end{gather}
where the tensor $\mathcal{C}_{pq}{}^{ijkl}$ is defined similarly to
$\mathcal{A}_m{}^{ijkl}$ and $\mathcal{B}_m{}^{ijkl}$ in
Eq.~(\ref{eq:C}). The two-forms $K_{mn}{}^{IJ}$ denote the derivative
of the Killing vectors $K_m{}^{IJ}$ on the round seven-sphere. 

During completion of this paper, a work by Oscar Varela derived
similar coordinate-free Ans\"atze for the metric, the warp factor and
the flux \cite{Varela:2015ywx}. These expressions however, are given
in a different form that is based on the tensor hierarchy formalism of
gauged supergravity \textbf{(}see Eqs.~(24-26) of
\cite{Varela:2015ywx}\textbf{)}. This makes it complicated to actually
compare my formulas to those of Varela's work. In order to illustrate
the simplicity of the Ans\"atze above, I test them for a G$_2$
invariant solution of 11-dimensional supergravity. This essential part
of the present work is done in Section \ref{sec:testing-new-uplift}.
It turns out that the new formulas in
Eqs.~(\ref{eq:metric-Intro}-\ref{eq:A3-Intro}) appear to be very
suitable for this test.

In the second part of this paper, \textbf{I derive a new uplift Ansatz
  for the internal four-form field-strength}
\begin{equation}
  \label{eq:F}
  F_{mnpq} = 4! \, \mathring D_{[m} A_{npq]}.
\end{equation}
Here, $\mathring D_m$ denotes the covariant derivative with respect to
the internal background metric $\mathring g_{mn}$. So far,
Eq.~(\ref{eq:F}) could only be used in particular cases --- when an
explicit expression for the internal three-form potential was already
given. However, it was rather complicated to compute the derivative of
$A_{mnp}$ in such cases, for example to find the G$_2$ or
SO(3)$\times$SO(3) invariant solutions of 11-dimensional supergravity
\cite{Godazgar:2013nma,Godazgar:2014eza}. With the new
\textit{general} Ansatz for $A_{mnp}$ above, I derive a simple direct
formula for the four-form field-strength, i.e.
\begin{equation}
  \label{eq:new-F-Ansatz-intro}
  F_{mnpq} = m_7 \Delta \mathring g_{s[m}
  \Big(
    4 
    \epsilon_{npq]r_1r_2r_3}{}^s A^{r_1r_2r_3} - 3
    g_{n|t|} A_{pq]r} K^{rs\,IJ} 
    K^{t\,KL}
    \left(
      u_{ij}{}^{IJ} + v_{ij\,IJ}
    \right)
    \left(
      u^{ij}{}_{KL} + v^{ij\,KL}
    \right)
  \Big).
\end{equation}
Here, $m_7$ denotes the inverse $S^7$ radius and $\epsilon_{r_1\cdots
  r_7}$ is the internal $\epsilon$-tensor. 

A formula for the complete four-form field-strength occurs in Eq.~(28)
of Varela's work \cite{Varela:2015ywx}. Again, it is hard to compare
both formulas because the expression in \cite{Varela:2015ywx} is given
in a form based on the tensor hierarchy formalism of gauged
supergravity. In Section \ref{sec:testing-new-uplift}, I will
demonstrate once more that the present Ansatz above is given in a very
convenient form --- it can be directly used for a test against the
G$_2$ invariant solution of 11-dimensional supergravity.

The new non-linear Ansatz in Eq.~(\ref{eq:new-F-Ansatz-intro})
provides another remarkable result: The above expression is `almost'
covariant\footnote{Indices of $\mathring g_{mn}$ and the Killing forms
  are raised and lowered with the background metric. All other tensors
  are covariant.}, which means that raising the indices is simple,
\begin{equation}
  \label{eq:new-F-raised-intro}
  F^{mnpq} = m_7 \Delta \mathring g_{st} g^{t[m}\Big( 4
  \epsilon^{npq]r_1r_2r_3s} A_{r_1r_2r_3} - 3 A^{np}{}_r K^{q]\,IJ}
  K^{rs\,KL}
  \left(
    u^{ij}{}_{IJ} + v^{ij\,IJ}
  \right)
  \left(
    u_{ij}{}^{KL} + v_{ij\,KL}
  \right) \Big).
\end{equation}
Up to now, it was far more complicated to derive $F^{mnpq}$ --- by
raising each single index of $F_{mnpq}$ with the explicit expression
for the inverse metric $g^{mn}$. For example, this was one of the
hardest tasks in verifying the SO(3)$\times$SO(3) invariant solution
of 11-dimensional supergravity \cite{Godazgar:2014eza}. In the case of
maximally symmetric spacetimes, these results can be used to compute
the components of the Ricci tensor via the equations of motion.


In the next section, I collect the main steps to find the consistent
uplift of $N=8$ supergravity to 11-dimensional supergravity. In
Section \ref{sec:strategy-finding-non}, I re-derive the known
non-linear Ans\"atze for the inverse metric $\Delta^{-1} g^{mn}$, the
three-form with mixed index structure $A_{mn}{}^p$ and the six-form
potential $A_{m_1\cdots m_6}$. In Section \ref{sec:non-linear-metric},
I present the new uplift Ans\"atze for the metric $g_{mn}$, the warp
factor $\Delta$ and the full internal three-form potential $A_{mnp}$.
Furthermore, I find the new non-linear Ansatz for the four-form
field-strength ($F_{mnpq}$ and $F^{mnpq}$) in Section
\ref{sec:non-linear-ansatz}. In Section \ref{sec:testing-new-uplift},
I test the new uplift Ans\"atze for the G$_2$ invariant solution of
11-dimensional supergravity: I compute the metric and the four-form
field-strength using the new formulas in
Eqs.~(\ref{eq:metric-Intro},\ref{eq:new-F-Ansatz-intro})\footnote{A
  combination of the old Ans\"atze for $\Delta^{-1}g^{mn}$,
  $A_{mn}{}^p$ and the new metric Ansatz yields the new formulas for
  the warp factor $\Delta$ and the internal three-form $A_{mnp}$. The
  old expressions for the inverse metric and the three-form potential
  with mixed index structure have already been tested in
  \cite{Godazgar:2013nma}. Hence, it suffices to test the Ans\"atze
  for the metric $g_{mn}$ and the four-form field-strength $F_{mnpq}$
  for a G$_2$ invariant solution of 11-dimensional supergravity.} and
compare with the results of \cite{Godazgar:2013nma}. Finally, I
conclude in Section \ref{sec:conclusion}.

\section{The uplift of $N=8$ supergravity to 11-dimensional
  supergravity}
\label{sec:preliminaries}

The bosonic field content of 11-dimensional supergravity is an elfbein
$E_M{}^A(x,y)$ and a three-form potential $A_{MNP}(x,y)$. The set of
coordinates splits into four spacetime (external) coordinates $x$ and
seven internal coordinates $y$. Capital Roman letters denote
11-dimensional indices. These split into external (Greek letters) and
internal indices (lower case Roman letters). As a rule of thumb:
Letters from the middle of an alphabet always denote curved spacetime
indices and letters from the beginning of an alphabet are the
corresponding tangent space indices.

The bosonic Lagrangian of 11-dimensional supergravity is written in
terms of the elfbein, the three-form potential and the four-form
field-strength \cite{Englert:1982vs}. The latter is defined by
\begin{equation}
  \label{eq:4form}
  F_{(4)} = \mathrm{d} A_{(3)} \qquad \Leftrightarrow \qquad F_{MNPQ}
  = 4!\, \partial_{[M} A_{NPQ]}.
\end{equation}
The Lagrangian can also be written in terms of dual fields
\cite{Nicolai:1980kb}: for example, one could replace $F_{(4)}$ by its
dual seven-form
\begin{equation}
  \label{eq:7form}
  F_{(7)} = \star F_{(4)}
\end{equation}
and the three-form potential by its dual six-form $A_{M_1\cdots M_6}$.
The latter is the potential for the dual seven-form field-strength,
\begin{equation}
  \label{eq:F7}
  F_{(7)} = \mathrm{d} A_{(6)} + 3 \sqrt{2} A_{(3)} \wedge F_{(4)} +
  \textrm{fermionic terms}.
\end{equation}
Later, one needs the six-form potential to describe certain vector and
scalar degrees of freedom.

Let us count the scalar and vector fields in 11-dimensional
supergravity. The elfbein is given by
\begin{equation}
  \label{eq:Elfbein}
  E_M{}^A =
  \begin{pmatrix}
    e_\mu{}^\alpha & B_\mu{}^m\, e_m{}^a \\ 0 & e_m{}^a
  \end{pmatrix}.
\end{equation}
It contains the vierbein $e_\mu{}^\alpha(x,y)$, seven vectors
$B_\mu{}^m(x,y)$ and 28 scalar fields $e_m{}^a(x,y)$. On the other
hand, the three-form potential splits into the components
\begin{equation}
  \label{eq:Three-form}
  A_{MNP} =
  \begin{pmatrix}
    A_{\mu\nu\rho}, & A_{\mu \nu m}, & A_{\mu mn}, & A_{mnp}
  \end{pmatrix}.
\end{equation}
There are 21 vector fields in $A_{\mu mn}(x,y)$. Furthermore,
$A_{\mu\nu m}(x,y)$ contains seven and $A_{mnp}(x,y)$ 35 scalar
degrees of freedom. The remaining components $A_{\mu\nu\rho}(x,y)$
represent the potential for the external field-strength
\begin{equation}
  \label{eq:external-4form}
  F_{\mu\nu\rho\sigma}(x,y) = 4! \,\partial_{[\mu}
  A_{\nu\rho\sigma]}(x,y) 
\end{equation}
and hence, contain no more scalar or vector degrees of freedom. This
is because for all dimensional reductions,
\begin{equation}
  \label{eq:Freund-Rubin}
  F_{\mu\nu\rho\sigma}(x,y) = i \frak f_\mathrm{FR}(x,y) \mathring
  \eta_{\mu \nu \rho \sigma}.
\end{equation}
The Freund-Rubin parameter $\frak f_\mathrm{FR}$ is constant for
Freund-Rubin compactifications \cite{Freund:1980xh} and $\mathring
\eta_{\mu \nu \rho \sigma}$ represents the volume form in four
dimensions. All in all, there are $7 + 21 = 28$ vectors and $28 + 7 +
35 = 70$ scalar degrees of freedom in 11-dimensional supergravity.

The bosonic field content of $N=8$ supergravity is a vierbein
$\mathring e_\mu{}^\alpha(x)$, 28 `electric' vector fields
$A_\mu{}^{IJ}(x)$ as well as 35 scalar and 35 pseudo-scalar fields
$u_{ij}{}^{IJ}(x)$, $v_{ij\,IJ}(x)$. All these fields only depend on
the four spacetime coordinates $x$. The (antisymmetric) bi-vector
indices $IJ$ belong to the 28-dimensional representation of
SL(8,$\mathbb{R}$) and the (antisymmetric) bi-vector indices $ij$
belong to the 28-dimensional representation of the local SU(8).
\textit{The bosonic degrees of freedom of both, $N=8$ supergravity and
  11-dimensional supergravity coincide. This is at least, necessary
  for a consistent uplift.}

In order to uplift $N=8$ supergravity to 11-dimensional supergravity,
one must explicitly relate the vierbeine, as well as the scalar and
vector fields of both theories to each other. In the following, I will
restrict to the $S^7$ compactification \cite{Duff:1983gq}. The
matching was found by comparing the supersymmetry transformations of
the four- and 11-dimensional fields
\cite{deWit:1984va,Godazgar:2013pfa}. It is based on a global
E$_{7(7)}$ symmetry in $N=8$ supergravity \cite{cremmerjulia}.
E$_{7(7)}$ is not a symmetry of 11-dimensional supergravity. However,
one may emphasize the respective E$_{7(7)}$ structures as much as
possible in order to compare the fields with those of $N=8$
supergravity.

The correct relation between the vierbeine of $N=8$ supergravity and
11-dimensional supergravity is
\begin{equation}
  \label{eq:uplift-vierbein}
  e_\mu{}^\alpha(x,y) = \Delta(x,y)^{-1/2} \mathring
  e_\mu{}^\alpha(x).
\end{equation}
The proportionality factor $\Delta(x,y)$ is called the warp factor.
Let $\mathring e_m{}^a$ be the siebenbein for the round seven-sphere
and $\mathring g_{mn}$ denote the respective background metric and let
$g_{mn}$ be the full internal metric of the deformed $S^7$
\cite{dWN13},
\begin{equation}
  \label{eq:7-metric}
  \mathring g_{mn} = \mathring e_m{}^a \mathring e_{n\,a}, \qquad
  g_{mn} = e_m{}^a e_{n\,a}.
\end{equation}
Then, the warp factor is defined by
\begin{equation}
  \label{eq:warp}
  \Delta = \frac{\det \left(e_m{}^a\right)}{\det \left( \mathring
      e_m{}^a\right)} = \sqrt{\frac{\det(g_{mn})}{\det\left(\mathring
        g_{mn}\right)}}. 
\end{equation}

In order to match the scalar degrees of freedom, one first observes
that the 35 scalars and 35 pseudo-scalars of $N=8$ supergravity
parametrize an element of E$_7/$SU(8). This co-set space is indeed,
70-dimensional. Both, scalars and pseudo-scalars together form an
element $\hat{\mathcal{V}}^\mathcal{M}{}_{ij}(x)$ in the fundamental
representation $\mathbf{56}$ of E$_{7(7)}$. Its SL(8,$\mathbb{R}$)
decomposition is given by
\begin{gather} 
  \label{eq:4d-scalars}
  \hat{\mathcal{V}}^\mathcal{M}{}_{ij} =
  \begin{pmatrix}
    \frac{i}{\sqrt{2}} \left( u_{ij}{}^{IJ} + v_{ij\,IJ} \right), & -
    \frac{1}{\sqrt{2}} \left( u_{ij}{}^{IJ} - v_{ij\,IJ} \right)
  \end{pmatrix},\\
  \label{eq:SL8-decomposition}
  \mathbf{56} \to \mathbf{28} \oplus 
  \overline{\mathbf{28}}.
\end{gather}
The $\mathbf{56}$ representation is labeled by indices
$\mathcal{M},\mathcal{N},\ldots$, which are raised and lowered with
the symplectic form $\Omega_{\mathcal{M}\mathcal{N}}$ (see
\cite{cremmerjulia}). The SU(8) indices $ij$ are raised and lowered
via complex conjugation,
\begin{equation}
  \label{eq:complex-conjugation}
  u^{ij}{}_{IJ} =
  \left(
    u_{ij}{}^{IJ}
  \right)^*, \qquad v^{ij\,IJ} =
  \left(
    v_{ij\,IJ}
  \right)^*.
\end{equation}

One also writes the scalar fields of 11-dimensional supergravity in an
E$_{7(7)}$ covariant way. Therefore, it is convenient to describe all
scalars by the fields $e_m{}^a$, $A_{m_1\cdots m_6}$ and $A_{mnp}$
(rather than using $A_{\mu\nu m}$). Indeed, the internal dual six-form
potential $A_{m_1\cdots m_6}$ contains the same scalar degrees of
freedom as $A_{\mu\nu m}$. In a second step, one converts this scalar
field content ($e_m{}^a$, $A_{m_1\cdots m_6}$ and $A_{mnp}$) into
components of a `56-bein' of E$_{7(7)}$, i.e.
\cite{Godazgar:2013dma,Godazgar:2014sla}
\begin{align}
  \mathcal{V}^{m}{}_{AB} =& - \frac{\sqrt{2}}{8} \Delta^{-1/2}
  \Gamma^m_{AB}, \label{eq:V11d1} \\
  \label{eq:V11d2}
  \mathcal{V}_{mn\;AB} =& - \frac{\sqrt{2}}{8} \Delta^{-1/2} \left(
    \Gamma_{mn\;AB} + 6\sqrt{2} A_{mnp} \Gamma^p_{AB}
  \right),\\
  \label{eq:V11d3}
  \mathcal{V}^{mn}{}_{AB} =& - \frac{\sqrt{2}}{8} \cdot \frac{1}{5!}
  \mathring \eta^{mnp_1\cdots p_5} \Delta^{-1/2} \Bigg[
  \Gamma_{p_1\cdots p_5\;AB} + 60\sqrt{2} A_{p_1p_2p_3} \Gamma_{p_4
    p_5\;AB} \nonumber\\ &\qquad\qquad\qquad\qquad\qquad\qquad -
  6!\sqrt{2} \left( A_{qp_1\cdots p_5} - \frac{\sqrt{2}}{4}
    A_{qp_1p_2} A_{p_3p_4p_5} \right) \Gamma^q_{AB}
  \Bigg],\\
  \mathcal{V}_{m\;AB} =& - \frac{\sqrt{2}}{8} \cdot \frac{1}{7!}
  \mathring \eta^{p_1 \cdots p_7} \Delta^{-1/2} \Bigg[
  (\Gamma_{p_1\cdots p_7}\Gamma_m)_{AB} + 126\sqrt{2}
  A_{mp_1p_2}\Gamma_{p_3\cdots p_7\;AB} \nonumber\\
  &\qquad\qquad\qquad\qquad\qquad\qquad + 3\sqrt{2}\cdot 7! \left(
    A_{mp_1\cdots p_5} + \frac{\sqrt{2}}{4} A_{mp_1p_2}A_{p_3p_4p_5}
  \right) \Gamma_{p_6p_7\;AB} \nonumber\\
  &\qquad\qquad\qquad\qquad\qquad\qquad + \frac{9!}{2} \left(
    A_{mp_1\cdots p_5} + \frac{\sqrt{2}}{12}A_{mp_1p_2}A_{p_3p_4p_5}
  \right) A_{p_6p_7q} \Gamma^q_{AB} \Bigg]. \label{eq:V11d4}
\end{align}
These components constitute the GL(7,$\mathbb{R}$) decomposition of
the 56-bein
\begin{gather}
  \label{eq:11d-scalars}
  \mathcal{V}^\mathcal{M}{}_{AB} =
  \begin{pmatrix}
    \mathcal{V}^m{}_{AB},& \mathcal{V}_{mn\,AB},&
    \mathcal{V}^{mn}{}_{AB},& \mathcal{V}_{m\,AB}
  \end{pmatrix},\\
  \label{eq:GL7-decomposition}
  \mathbf{56} \to \mathbf{7} \oplus \mathbf{21} \oplus
  \overline{\mathbf{21}} \oplus \overline{\mathbf{7}}.
\end{gather}
The SU(8) indices $A,B,\ldots$ are raised and lowered by complex
conjugation\footnote{It should always be clear from the context
  whether $A,B,\ldots$ are SU(8)- or 11-dimensional tangent space
  indices.} and the 8$\times$8 $\Gamma$-matrices are defined in
Appendix \ref{sec:kill-spin-kill}.

The correct relation between the 56-bein in 11 dimensions and the
four-dimensional scalars $\hat{\mathcal{V}}$ of $N=8$ supergravity was
found by considering the respective supersymmetry transformations
\cite{Godazgar:2013pfa}\footnote{Note that initially,
  Eq.~(\ref{eq:uplift-scalars}) follows from the respective uplift
  relation for the vectors in Eq.~(\ref{eq:uplift-vector}).}. It is
given by
\begin{equation} \label{eq:uplift-scalars}
  \mathcal{V}^{\mathcal{M}}{}_{AB}(x,y) =
  \mathcal{R}^{\mathcal{M}}{}_{\mathcal{N}}(y)\, \eta_A^{i}(y)\,
  \eta_B^j(y)\, \hat{\mathcal{V}}^{\mathcal{N}}{}_{ij}(x).
\end{equation}
Here, $\eta_A^{i}$ are the eight Killing spinors defined on the
internal geometry. The upper index $\mathcal{M}$ of the transformation matrix
$\mathcal{R}^\mathcal{M}{}_\mathcal{N}$ is decomposed under
GL(7,$\mathbb{R}$)
\textbf{(}Eq.~(\ref{eq:GL7-decomposition})\textbf{)} whereas the lower
index $\mathcal{N}$ is decomposed under SL(8,$\mathbb{R}$)
\textbf{(}Eq.~(\ref{eq:SL8-decomposition})\textbf{)},
\begin{equation}
  \label{eq:R-components}
  \mathcal{R}^\mathcal{M}{}_\mathcal{N} =
  \begin{pmatrix}
    \mathcal{R}^m{}_{IJ} & \mathcal{R}^{m\,IJ}\\
    \mathcal{R}_{mn\,IJ} & \mathcal{R}_{mn}{}^{IJ}\\[2.5pt]
    \mathcal{R}^{mn}{}_{IJ} & \mathcal{R}^{mn\,IJ}\\
    \mathcal{R}_{m\,IJ} & \mathcal{R}_{m}{}^{IJ}
  \end{pmatrix}.
\end{equation}
The non-zero components are \cite{Godazgar:2013pfa}
\begin{align}
  \label{eq:R-explicitA}
  \mathcal{R}^m{}_{IJ}(y) =& \,\frac{1}{4} K^{m\,IJ}(y),\\
  \mathcal{R}_{mn}{}^{IJ}(y) =& \,\frac{1}{4} K_{mn}{}^{IJ}(y),\\
  \mathcal{R}^{mn}{}_{IJ}(y) =& \,\frac{1}{4} \left( 2
    \mathring \zeta^{[m} K^{n]\,IJ} - K^{mn\,IJ}
  \right) (y),\\
  \label{eq:R-explicitD}
  \mathcal{R}_m{}^{IJ}(y) =& \,\frac{1}{4} \left(
    \mathring \zeta^n K_{mn}{}^{IJ} - K_m{}^{IJ} \right) (y).
\end{align}
They depend on the Killing vectors $K_m{}^{IJ}(y)$ and -forms
$K_{mn}{}^{IJ}(y)$ as well as on the dual volume potential $\mathring
\zeta^m(y)$ of the seven-sphere. The Killing vectors and -forms are
defined in Appendix \ref{sec:kill-spin-kill}. The (seven dimensional)
dual of $\mathring \zeta^m(y)$ is the six-form potential for the
internal background volume form $\mathring \eta_{m_1\cdots m_7}$,
\begin{gather}
  \label{eq:*zeta}
  \mathring \zeta^n = 6 \, \mathring \eta^{n m_1\cdots m_6}
  \mathring \zeta_{m_1 \cdots m_6}, \qquad
  \mathring \zeta_{m_1\cdots m_6} = \frac{1}{6\cdot 6!}\, \mathring
  \eta_{m_1\cdots m_7} \, \mathring \zeta^{m_7},\\
  \label{eq:zeta}
  7! \mathring D_{[m_1} \mathring \zeta_{m_2\cdots m_7]} = m_7
  \mathring \eta_{m_1\cdots m_7}.
\end{gather}
Note the non-standard normalization of $\mathring \zeta^m$, which is
more convenient for my purposes. $m_7$ denotes the inverse radius of
the round $S^7$.

Using
Eqs.~(\ref{eq:4d-scalars},\ref{eq:R-explicitA}-\ref{eq:R-explicitD}),
one finally finds the components of
\begin{equation}
  \label{eq:4d-scalars-nohat}
  \mathcal{V}^\mathcal{M}_{ij}(x,y) =
  \mathcal{R}^{\mathcal{M}}{}_{\mathcal{N}}(y) 
  \hat{\mathcal{V}}^{\mathcal{N}}{}_{ij}(x),
\end{equation}
namely
\begin{gather}
  \label{eq:4d-vielbeinB}
  \mathcal{V}^{m8}{}_{ij}(x,y) = \frac{\sqrt{2}i}{8} K^{m\;IJ}(y)
  \left( u_{ij}{}^{IJ} + v_{ij\;IJ}
  \right)(x),\\ \label{eq:4d-vielbeinC} \mathcal{V}_{mn\;ij}(x,y)=
  -\frac{\sqrt{2}}{8} K_{mn}{}^{IJ}(y) \left( u_{ij}{}^{IJ} -
    v_{ij\;IJ}
  \right)(x),\\ \label{eq:4d-vielbeinD}
  \mathcal{V}^{mn}{}_{ij} (x,y)= \frac{\sqrt{2}i}{8} \left( 2
    \mathring \zeta^{[m} K^{n]\,IJ} - K^{mn\,IJ} \right)(y) \left(
    u_{ij}{}^{IJ} + v_{ij\;IJ}
  \right)(x),\\
  \mathcal{V}_{m8\;ij} (x,y) = -\frac{\sqrt{2}}{8} \left( \mathring
    \zeta^n K_{mn}{}^{IJ} - K_m{}^{IJ} \right)(y) \left( u_{ij}{}^{IJ}
    - v_{ij\;IJ} \right)(x).
  \label{eq:4d-vielbeinE}
\end{gather}

In order to match the vector degrees of freedom, one first dualizes
the 28 `electric' vector fields $A_\mu{}^{IJ}(x)$ in $N=8$
supergravity to form 28 `magnetic' vector fields $A_\mu{}_{IJ}(x)$.
Only electric and magnetic vector fields together fit into the
$\mathbf{56}$ representation of E$_{7(7)}$: they represent the
SL(8,$\mathbb{R}$) decomposition of
\begin{gather}
  \label{eq:4d-electric-magnetic-vf}
  A_\mu{}^\mathcal{M} =
  \left(
    A_\mu{}^{IJ}, A_\mu{}_{IJ}
  \right)
\end{gather}
along the lines of Eq.~(\ref{eq:SL8-decomposition}). One also extends
the 28 vector fields $B_\mu{}^m$ and $A_{\mu mn}$ in 11-dimensional
supergravity such that they fit into the $\mathbf{56}$ representation
of E$_{7(7)}$. There are 21 dual vectors $A_{\mu m_1\cdots m_5}$
coming from the six-form potential and seven `dual graviphotons' that
have no physical interpretation \cite{Godazgar:2013dma}. Similar to
the case of scalar fields, one defines a 56-bein $B_\mu{}^\mathcal{M}$
of E$_{7(7)}$, which decomposes under GL(7,$\mathbb{R}$) into the
various vector degrees of freedom above. Since this work concentrates
on the uplift of the scalar fields, I do not give the explicit
GL(7,$\mathbb{R}$) decomposition for $B_\mu{}^\mathcal{M}$ here. The
interested reader may have a look at
\cite{Godazgar:2013dma,Godazgar:2013pfa,Godazgar:2014sla}.

The consistent relation between the vector fields
$A_\mu{}^\mathcal{M}(x)$ of $N=8$ supergravity and the 11-dimensional
vectors $B_\mu{}^\mathcal{M}(x,y)$ is similar to
Eq.~(\ref{eq:uplift-scalars})\footnote{The last seven components of
  $B_\mu{}^\mathcal{M}$ belong to the non-physical dual graviphotons.
  Eq.~(\ref{eq:uplift-vector}) therefore, does only make sense in the
  first 49 components.},
\begin{equation}
  \label{eq:uplift-vector}
  B_\mu{}^\mathcal{M}(x,y) = \mathcal{R}^\mathcal{M}{}_\mathcal{N}(y)
  A_\mu{}^\mathcal{N}(x).
\end{equation}
It has also been found by a careful analysis of the supersymmetry
transformations in four and 11 dimensions.

Here is a simple example for the readers convenience: The first seven
components of $B_\mu{}^\mathcal{M}$ are proportional to the vectors
$B_\mu{}^m$. With Eqs.~(\ref{eq:uplift-vector},\ref{eq:R-explicitA})
one then finds the old Ansatz for the vector fields in Kaluza-Klein
theory \cite{Witten:1981me}, i.e.
\begin{equation}
  \label{eq:Example-vectors}
  B_\mu{}^m(x,y) \propto K^{m\,IJ}(y) A_\mu{}^{IJ}(x).
\end{equation}

The task of uplifting $N=8$ supergravity to 11-dimensional
supergravity is now the following: Starting from
Eqs.~(\ref{eq:uplift-scalars},\ref{eq:uplift-vector}), one must seek
explicit expressions for the 11-dimensional vector and scalar fields
in terms of the four-dimensional ones,
\begin{align}
  \label{eq:uplift-needs}
  \begin{pmatrix}
    B_\mu{}^m,& A_{\mu mn},& A_{\mu m_1\cdots m_5},& \textrm{dual
      graviphotons}
  \end{pmatrix} \qquad  \Leftrightarrow& \qquad
  \begin{pmatrix}
    A_\mu{}^{IJ},& A_{\mu\,IJ}
  \end{pmatrix},\\[5pt]
  \begin{pmatrix}
    g_{mn}, & A_{mnp}, & A_{m_1\cdots m_6}
  \end{pmatrix} \qquad\qquad\quad\quad \Leftrightarrow& \qquad
  \begin{pmatrix}
    u_{ij}{}^{IJ}, & v_{ij\,IJ}
  \end{pmatrix}.
\end{align}
In principle, these relations have been found in
\cite{deWit:1984nz,Godazgar:2013pfa}. However, instead of a relation
for the metric $g_{mn}(x,y)$, the authors only found an expression for
the inverse metric $\Delta^{-1} g^{mn}(x,y)$, scaled with the warp
factor. Furthermore, the Ans\"atze for the three-form and six-form
potentials require the full metric $g_{mn}$. Until now, the inversion
of $\Delta^{-1}g^{mn}$ is only possible in particular cases, e.g. for
G$_2$, SO(3)$\times$SO(3) or SU(3)$\times$U(1)$\times$U(1) invariant
solutions
\cite{Godazgar:2013nma,Godazgar:2014eza,Pilch:2015vha,Pilch:2015dwa}.
Also the warp factor can only be computed from an explicit expression
for the metric $g_{mn}$ (by taking the determinant).

The reader familiar with the uplift Ans\"atze presented in
\cite{Godazgar:2013pfa} may skip the next section, which repeats the
derivation of the known scalar uplifts. Section
\ref{sec:non-linear-metric} then presents new non-linear Ans\"atze for
the full internal metric $g_{mn}$, the warp factor $\Delta$ and the
internal three-form potential $A_{mnp}$. These hold for the uplift of
$N=8$ supergravity to 11-dimensional maximally gauged supergravity,
even without further restrictions \textbf{(}such as G$_2$,
SO(3)$\times$SO(3) or SU(3)$\times$U(1)$\times$U(1)
invariance\textbf{)}.

\section{Known Ans\"atze for $\Delta^{-1}g^{mn}$, $A_{mn}{}^p$ and
  $A_{m_1\cdots m_6}$}
\label{sec:strategy-finding-non}

For the readers convenience, I repeat the steps to derive the known
uplift relations for the inverse metric $\Delta^{-1}g^{mn}$, the
three-form with mixed index structure $A_{mn}{}^p$ and the six-form
potential $A_{m_1\cdots m_6}$. This was done in
\cite{Godazgar:2013pfa} and is the basis to understand the new
Ans\"atze for the metric $g_{mn}$, the warp factor $\Delta$ and the
full internal three-form potential $A_{mnp}$ in Section
\ref{sec:non-linear-metric}.

The main problem of comparing the vielbein components in
Eqs.~(\ref{eq:V11d1}-\ref{eq:V11d4}) and
Eqs.~(\ref{eq:4d-vielbeinB}-\ref{eq:4d-vielbeinE}) is the occurrence of
the Killing spinors in Eq.~(\ref{eq:uplift-scalars}). However, these
are orthonormal and would drop out in non-linear SU(8)-invariant
combinations of the vielbeine. For example, let us consider the
expression
\begin{equation}
  \label{eq:Ansatz-inverse-metric}
  \mathcal{V}^{m}{}_{AB} \mathcal{V}^{n}{}^{AB} = \eta^i_A \eta^j_B
  \mathcal{V}^m{}_{ij} \eta^A_k \eta^B_l \mathcal{V}^{n\,kl} = 
  \mathcal{V}^{m}{}_{ij} \mathcal{V}^{n}{}^{ij}.   
\end{equation}
Indeed, the Killing spinors $\eta_A^i(y)$ drop out. One now uses
Eq.~(\ref{eq:V11d1}) on the lhs and Eq.~(\ref{eq:4d-vielbeinB}) on the
rhs, which results in a non-linear uplift Ansatz for the inverse
metric, i.e.
\begin{equation}
  \label{eq:inverse-metric}
  \Delta^{-1} g^{mn}(x,y) = \frac{1}{8} K^{m\;IJ}(y) K^{n\;KL}(y)
  \left( u_{ij}{}^{IJ} + v_{ij\;IJ}
  \right)(x) \left( u^{ij}{}_{KL} + v^{ij\;KL}
  \right)(x).
\end{equation}
Here, I used the Clifford algebra of the $\Gamma$-matrices, given in
Appendix \ref{sec:kill-spin-kill}.

In a similar way, one relates
\begin{equation}
  \label{eq:Ansatz-mixed-3form}
  \mathcal{V}_{mn}{}^{AB} \mathcal{V}^{p8}{}_{AB} =
  \mathcal{V}_{mn}{}^{ij} \mathcal{V}^{p8}{}_{ij},
\end{equation}
which yields a non-linear uplift Ansatz for the three-form. Indeed,
using Eqs.~(\ref{eq:V11d1},\ref{eq:V11d2}) on the lhs as well as
Eqs.~(\ref{eq:4d-vielbeinB},\ref{eq:4d-vielbeinC}) on the rhs, one
finds
\begin{equation}
  \label{eq:mixed-3-form}
  \Delta^{-1} A_{mn}{}^p(x,y) = -\frac{\sqrt{2}i}{96} K_{mn}{}^{IJ}(y)
  K^{p\;KL}(y) 
  \left( u^{ij}{}_{IJ} -
    v^{ij\;IJ}
  \right)
  \left( u_{ij}{}^{KL} + v_{ij\;KL}  \right) (x).
\end{equation}

In order to derive an uplift Ansatz for the internal six-form
potential $A_{m_1\cdots m_6}$, I introduce the (seven dimensional)
dual one-form
\begin{equation}
  \label{eq:*A6}
  A^n = 6\, \epsilon^{nm_1\cdots m_6} A_{m_1\cdots m_6}.
\end{equation}
Similar to the dual volume potential on the round seven-sphere,
$\mathring \zeta^m$, I use a non-standard normalization for later
convenience. The six-form potential $A_{(6)}$ is a tensor in the
internal space and its (seven dimensional) dual $A_{(1)}$ is
constructed with the full $\epsilon$-tensor. However, one can convert
this $\epsilon$-tensor to the tensor density $\mathring \eta$ $(= \pm
1,0)$ using the internal seven-bein $e_m{}^a$
\begin{equation}
  \label{eq:tensor-density}
  \epsilon_{m_1\cdots m_7} = e_{m_1}{}^{a_1}\ldots e_{m_7}{}^{a_7}
  \mathring \eta_{a_1\cdots a_7} = \Delta \mathring \eta_{m_1\cdots m_7}.
\end{equation}
Here, I used the definition of the warp factor in Eq.~(\ref{eq:warp}).
Eq.~(\ref{eq:*A6}) then reads
\begin{equation}
  \label{eq:*A1}
  A^n = \frac{6}{ \Delta} \mathring \eta^{nm_1\cdots m_6}
  A_{m_1\cdots m_6} \qquad \Leftrightarrow \qquad A_{m_1\cdots m_6} =
  \frac{\Delta}{6\cdot 6!} \mathring \eta_{m_1\cdots m_7} A^{m_7}.
\end{equation}
Note that the indices of the six-form potential and its dual are raised
and lowered with the full internal metric.

Now, let us consider the relation
\begin{equation}
  \mathcal{V}^{mn}{}_{AB} \mathcal{V}^{p8}{}^{AB} =
  \mathcal{V}^{mn}{}_{ij} \mathcal{V}^{p8}{}^{ij}
\end{equation}
and insert the various vielbein components in
Eqs.~(\ref{eq:V11d1},\ref{eq:V11d3}) and
Eqs.~(\ref{eq:4d-vielbeinB},\ref{eq:4d-vielbeinD}). This gives an
equation for $A^n$, i.e.
\begin{equation}
  \label{eq:A1-ansatz}
  \frac{\sqrt{2}}{9}
  \left(
    \Delta A^{[m} + 3\sqrt{2} \mathring \zeta^{[m}
  \right) g^{n]p} = \mathring \eta^{mn q_1\cdots q_5} A^p{}_{q_1q_2}
  A_{q_3 q_4 q_5} + \frac{\Delta}{24} K^{mn\,IJ} K^{p\,KL}
  \left(
    u_{ij}{}^{IJ} + v_{ij\,IJ}
  \right)
  \left(
    u^{ij}{}_{KL} + v^{ij\,KL}
  \right).
\end{equation}
When contracting this relation with $g_{np}$, the first term on the
rhs drops out because
\begin{equation}
  A_{[mnp}A_{qrs]} = 0.
\end{equation}
One finds
\begin{equation}
  \label{eq:A1-ansatz-contracted}
  \Delta A^{m}(x,y) + 3\sqrt{2} \mathring \zeta^{m}(y) =
  \frac{\Delta(x,y)}{8\sqrt{2}} g_{np}(x,y) K^{mn\,IJ}(y) K^{p\,KL}
  (y) 
  \left(
    u_{ij}{}^{IJ} + v_{ij\,IJ}
  \right)
  \left(
    u^{ij}{}_{KL} + v^{ij\,KL}
  \right)(x)
\end{equation}
and dualizes this expression using Eq.~(\ref{eq:*A1},\ref{eq:*zeta}),
\begin{equation}
  \label{eq:A6-ansatz}
  A_{m_1\cdots m_6} + 3\sqrt{2} \mathring \zeta_{m_1\cdots
    m_6} = \frac{\sqrt{2}}{96 \cdot 6!} \, \epsilon_{n m_1\cdots m_6}
  \, g_{pq} K^{np\,IJ} K^{q\,KL}
  \left(
    u_{ij}{}^{IJ} + v_{ij\,IJ}
  \right)
  \left(
    u^{ij}{}_{KL} + v^{ij\,KL}
  \right).
\end{equation}
Here, I suppressed the explicit dependence on the coordinates.

The rhs of Eqs.~(\ref{eq:A1-ansatz-contracted},\ref{eq:A6-ansatz})
further simplifies using the uplift Ansatz for the inverse metric in
Eq.~(\ref{eq:inverse-metric}) and the definition of the Killing
two-form in Eq.~(\ref{eq:Killing-derivatives}). It is proportional to
\begin{equation}
  \label{eq:DlogD}
  \mathring D^{m} \log \Delta = \Delta^{-1} \mathring D^{m} \Delta =
  \frac{1}{2} g^{pq} \mathring D^m g_{pq},
\end{equation}
which finally gives a simpler non-linear Ansatz for the six-form
potential, i.e.
\begin{align}
  \label{eq:A1}
  \Delta A^m(x,y) + 3 \sqrt{2} \mathring \zeta^m(y) =&\,
  \frac{9\sqrt{2}}{4 m_7} \mathring D^m \log \Delta(x,y), \\ 
  \label{eq:A6}
  A_{m_1 \cdots m_6}(x,y) + 3\sqrt{2} \mathring \zeta_{m_1\cdots
    m_6}(y) =&\, \frac{\sqrt{2}}{16\cdot 5! m_7} \mathring \eta_{m_1
    \cdots m_7} \mathring D^{m_7} \log \Delta(x,y).
\end{align}
This result has already been derived in \cite{Godazgar:2015qia}. In
comparison to
Eqs.~(\ref{eq:A1-ansatz-contracted},\ref{eq:A6-ansatz}), the Ans\"atze
in Eqs.~(\ref{eq:A1},\ref{eq:A6}) do not contain the metric $g_{mn}$.
However, they require an explicit expression for the warp factor,
which also can only be given in particular cases.

\section{New non-linear Ans\"atze for the metric $g_{mn}$, the warp
  factor $\Delta$ and the full internal three-form potential
  $A_{mnp}$}
\label{sec:non-linear-metric}

In this section, I derive a new non-linear metric Ansatz for the
uplift of SO(8) gauged $N=8$ supergravity to 11-dimensional
supergravity. In combination with the expressions for the inverse
metric and the three-form with mixed index structure in
Eqs.~(\ref{eq:inverse-metric},\ref{eq:mixed-3-form}), I find further
uplift Ans\"atze for the warp factor and the internal three-form
potential $A_{mnp}$. Note that recent work derived similar
coordinate-free formulas \textbf{(}Eqs.~(24-26) of
\cite{Varela:2015ywx}\textbf{)} in a different form.

Following the strategy of the previous section, I consider the
relation
\begin{equation}
  \label{eq:pre-metric}
  \mathcal{V}_{mp}{}_{AB} \mathcal{V}^{p8}{}_{CD}
  \mathcal{V}_{nq}{}^{[AB} \mathcal{V}^{q8}{}^{CD]} =
  \mathcal{V}_{mp}{}_{ij} \mathcal{V}^{p8}{}_{kl}
  \mathcal{V}_{nq}{}^{[ij} \mathcal{V}^{q8}{}^{kl]}.
\end{equation}
Let us use Eqs.~(\ref{eq:V11d1},\ref{eq:V11d2}) on the lhs: All terms
including a factor of $A_{mnp}$ are of the form
\begin{equation}
  \label{eq:2}
  \ldots A_{mnp} \Gamma^n{}_{[AB} \Gamma^p{}_{CD]} \ldots = 0
\end{equation}
but such expressions vanish because an antisymmetric index pair $[np]$
is contracted with a symmetric index pair $(np)$. 

One finally computes the traces of the $\Gamma$-matrices using
Eq.~(\ref{eq:Gamma-prop6}) and finds that the lhs of
Eq.~(\ref{eq:pre-metric}) is proportional to the metric $g_{mn}$,
\begin{equation}
  \label{eq:pre-g}
  \Delta^{-2} g_{mn} = \frac{16}{3} \mathcal{V}_{mp\,ij}
  \mathcal{V}^p{}_{kl} \mathcal{V}_{nq}{}^{[ij} \mathcal{V}^{q\,kl]}.
\end{equation}
For the rhs, I use Eqs.~(\ref{eq:4d-vielbeinB},\ref{eq:4d-vielbeinC})
and find that
\begin{equation}
  \label{eq:pre-V}
  \mathcal{V}_{mp\,[ij} \mathcal{V}^p{}_{kl]} = -\frac{i}{32}
  K_{mp}{}^{IJ} K^{p\,KL}
  \left(
    u_{[ij}{}^{IJ} - v_{[ij\,IJ}
  \right)
  \left(
    u_{kl]}{}^{KL} + v_{kl]\,KL}
  \right).
\end{equation}
For some readers, Eqs.~(\ref{eq:pre-g},\ref{eq:pre-V}) together
already represent a useful metric Ansatz in terms of the Killing forms
and the four dimensional scalar fields. However, one may simplify the
resulting expression further: Using
Eqs.~(\ref{eq:K-prop1},\ref{eq:K-prop2}) in Appendix
\ref{sec:kill-spin-kill} yields
\begin{equation}
  \mathcal{V}_{mp}{}_{[ij} \mathcal{V}^p{}_{kl]} = -\frac{i}{8}
  \left(
    \mathcal{A}_m{}_{ijkl} - \mathcal{B}_m{}_{ijkl}
  \right),
\end{equation}
where I defined the convenient tensors
\begin{align}
  \label{eq:A}
  \mathcal{A}_m{}_{ijkl}(x,y) =&\, \frac{1}{4} K_{mn}{}^{[IJ}(y)
  K^n{}^{KL]}(y) \left( u_{ij}{}^{IJ} u_{kl}{}^{KL} - v_{ij\,IJ}
    v_{kl\,KL} \right) (x)
  ,\\
  \label{eq:B}
  \mathcal{B}_m{}_{ijkl}(x,y) =&\, K_m{}^{IJ}(y) \left( u_{ij}{}^{IK}
    v_{kl\,JK} - v_{ij\,IK} u_{kl}{}^{JK} \right) (x).
\end{align}
By definition, these are totally antisymmetric in the SU(8) indices
$[ijkl]$ and depend on all 11 coordinates $(x,y)$. Note that a certain
linear combination of both tensors is equal to the `non-metricity'
$\mathcal{P}_{m\,ijkl}$ in the SO(8) invariant vacuum
\cite{deWit:1986iy,Godazgar:2015qia} \footnote{In \cite{deWit:1986iy},
  the non-metricity $\mathcal{P}_{m\,ijkl}$ was denoted by
  $\mathcal{A}_{m\,ijkl}$.}. One finally finds the metric Ansatz in
terms of these tensors, i.e.
\begin{equation}
  \label{eq:metric}
  \Delta^{-2} g_{mn}(x,y) = \,\frac{1}{12}
  \left( \mathcal{A}_m{}_{ijkl} -
    \mathcal{B}_m{}_{ijkl} 
  \right)
  \left(
    \mathcal{A}_n{}^{ijkl} -
    \mathcal{B}_n{}^{ijkl}
  \right)(x,y).
\end{equation}
This Ansatz is quartic in the four-dimensional scalar fields
$u_{ij}{}^{IJ}$ and $v_{ij\,IJ}$, whereas the Ans\"atze for the
inverse metric and the mixed three-form potential were only quadratic.

Let us combine the Ans\"atze for the metric and the inverse metric in
Eqs.~(\ref{eq:metric},\ref{eq:inverse-metric}) to get a new Ansatz for
the warp factor. This can be done because the new metric Ansatz
contains a proportionality factor of $\Delta^{-2}$. One finds
\begin{equation}
  \label{eq:warp-Ansatz}
  \Delta^{-3}(x,y) = \frac{1}{28 \cdot 4!}
  \,\mathcal{C}_{ij}{}^{klmn}(x,y) \mathcal{C}^{ij}{}_{klmn}(x,y),
\end{equation}
where the tensor $\mathcal{C}_{pq}{}^{ijkl}$ is defined as
\begin{align}
  \label{eq:C}
  \mathcal{C}_{pq}{}^{ijkl}(x,y) =& K^{m}{}^{IJ}(y) \left(
    u_{pq}{}^{IJ} + v_{pq\;IJ} \right)(x) \left( \mathcal{A}_m{}^{ijkl} -
    \mathcal{B}_m{}^{ijkl} \right) (x,y).
\end{align}

Similarly, one combines the Ansatz for the three-form with mixed index
structure in Eq.~(\ref{eq:mixed-3-form}) with the metric Ansatz in
Eq.~(\ref{eq:metric}) to obtain a new Ansatz for the full internal
three-form potential, i.e.
\begin{equation}
  \label{eq:A-down}
  \Delta^{-3} A_{mnp}(x,y) = - \frac{\sqrt{2}i}{48\cdot 4!} K_{mn}{}^{IJ}(y)
  \left( u^{ij}{}_{IJ} -
    v^{ij\;IJ}
  \right)(x) \, \mathcal{C}_{ij}{}^{qrst} (x,y) \left( \mathcal{A}_{p\,qrst} -
    \mathcal{B}_{p\,qrst}\right) (x,y).
\end{equation}
The new Ans\"atze for the warp factor and the three-form potential are
sextic in the scalar fields $u_{ij}{}^{IJ}$ and $v_{ij\,IJ}$.

It may still be possible to simplify the new Ans\"atze using some
E$_{7(7)}$ properties of the $u_{ij}{}^{IJ}$ and $v_{ij\,IJ}$ tensors
\cite{deWit:1982ig,deWit:1986iy}. One such simplification concerns the
$C_{pq}{}^{ijkl}$ tensor that occurs in both, the warp factor and the
three-form potential. For the rest of this section, I show that it
factorizes into\footnote{I thank Hadi Godazgar for pointing this out.}
\begin{equation}
  \label{eq:C-delta}
  \mathcal{C}_{pq}{}^{ijkl}(x,y) = \frac{4}{3} \delta^{[i}{}_{[p}
  \left(
    \mathcal{C}_{1\,q]}{}^{jkl]}(x,y) + 2
    \mathcal{C}_{2\,q]}{}^{jkl]}(x,y) - 2 T_{q]}{}^{jkl]}(x)
  \right),
\end{equation}
where 
\begin{align}
  \label{eq:C1}
  \mathcal{C}_{1\,p}{}^{ijk}(x,y) =&\, K^{IJKL}(y) \left(
    u^{jk}{}_{IJ} + v^{jk\,IJ} \right) \left( u^{im}{}_{KM}
    u_{pm}{}^{LM} - v^{im\,KM}
    v_{pm\,LM} \right)(x),\\
  \mathcal{C}_{2\,p}{}^{ijk}(x,y) =&\, K^{IJKL}(y) \left(
    u^{jk}{}_{IM} + v^{jk\,IM} \right) \Big[ \left( u^{im}{}_{[JK}
    v_{pm\,LM]} - v^{im\,[JK} u_{pm}{}^{LM]}
  \right) \nonumber\\
  & \qquad \qquad \qquad \qquad \qquad \qquad- \frac{1}{8} \delta^i_p
  \left( u^{mn}{}_{[JK} v_{mn\,LM]} - v^{mn\,[JK} u_{mn}{}^{LM]}
  \right) \Big](x).
  \label{eq:C2}
\end{align}

The selfdual tensor $K^{IJKL}$ is defined as a certain combination of
Killing vectors in Eq.~(\ref{eq:K4}). It satisfies some useful
relations given in Appendix \ref{sec:kill-spin-kill}. The third term
in Eq.~(\ref{eq:C-delta}) represents the $T$-tensor, which is defined
in \cite{deWit:1982ig},
\begin{equation}
  \label{eq:T-tensor}
  T_i{}^{jkl}(x) =
  \left(
    u^{kl}{}_{IJ} + v^{kl\,IJ}
  \right)
  \left(
    u_{im}{}^{JK} u^{jm}{}_{KI} - v_{im\,JK} v^{jm\,KI}
  \right)(x).
\end{equation}
It only depends on spacetime coordinates $x$ and satisfies the
property
\begin{equation}
  \label{eq:T-tensor1}
  \left(
    u_{pq}{}^{IJ} + v_{pq\,IJ}
  \right)
  \left(
    u^{ij}{}_{IK} v^{kl\,JK} - v^{ij\,IK} u^{kl}{}_{JK}
  \right) = \frac{4}{3} \delta^{[i}{}_{[p} T_{q]}{}^{jkl]}.
\end{equation}
For further relations concerning the $T$-tensor, see
\cite{deWit:1982ig,deWit:1986iy}. Note that the only difference
between $C_{1\,p}{}^{ijk}$ and the $T$-tensor is the $K^{IJKL}$-factor
in Eq.~(\ref{eq:C1}) instead of a $\delta^{IJ}_{KL}$-factor in
Eq.~(\ref{eq:T-tensor}). This gives rise to interpret $\mathcal{C}_1$
and $\mathcal{C}_2$ as the \textit{$y$-dependent twins of the
  $T$-tensor.}

In order to prove Eq.~(\ref{eq:C-delta}), one starts with
Eq.~(\ref{eq:C}) and replaces the tensors $\mathcal{A}_m{}^{ijkl}$ and
$\mathcal{B}_m{}^{ijkl}$ with the respective expressions in
Eqs.~(\ref{eq:A},\ref{eq:B}). Secondly, using
Eqs.~(\ref{eq:K-prop0},\ref{eq:K-Identitites3}) gives
\begin{align}
  \label{eq:C2}
  \mathcal{C}_{pq}{}^{ijkl} =& -2 K^{IJKL}
  \left(
    u_{pq}{}^{IM} + v_{pq\,IM}
  \right)
  \left(
    u^{ij}{}_{[JK} u^{kl}{}_{LM]} - v^{ij\,[JK} v^{kl\,LM]}
  \right) \nonumber\\
  &- K^{IJKL}
  \left(
    u_{pq}{}^{KL} + v_{pq\,KL}
  \right)
  \left(
    u^{ij}{}_{IM} v^{kl\,JM} - v^{ij\,IM} u^{kl}{}_{JM}
  \right) - \frac{8}{3} \delta^{[i}{}_{[p} T_{q]}{}^{jkl]},
\end{align}
which can be rearranged,
\begin{align}
  \label{eq:C3}
  \mathcal{C}_{pq}{}^{ijkl} =&\, 2 K^{IJKL} \left( u^{[ij}{}_{IJ} +
    v^{[ij\,IJ} \right) \left( u^{kl]}{}_{KM} u_{pq}{}^{LM} -
    v^{kl]\,KM} v_{pq\,LM} \right) \nonumber\\
  &+ 4 K^{IJKL} \left( u^{[ij}{}_{IM}
    + v^{[ij\,IM} \right) \left( u^{kl]}{}_{[JK} v_{pq\,LM]} -
    v^{kl]\,[JK} u_{pq}{}^{LM]} \right) - \frac{8}{3}
  \delta^{[i}{}_{[p} T_{q]}{}^{jkl]}.
\end{align}
Finally, I use Eq.~(4.7) of \cite{deWit:1982ig} and Eq.~(5.21) of
\cite{deWit:1986iy}, i.e.
\begin{align}
  \label{eq:u-v-equationsA}
  \left(u^{ij}{}_{IM} u_{kl}{}^{JM} - v^{ij\,IM} v_{kl\,JM}
  \right)\big|_{[IJ]} =&\, \frac{2}{3} \delta^{[i}{}_{[k}
  \left(u^{j]m}{}_{IM} u_{l]m}{}^{JM} - v^{j]m\,IM} v_{l]m\,JM}
  \right)\Big|_{[IJ]} \\
  \left( u^{ij}{}_{IJ} v_{kl\,KL} - v^{ij\,IJ} u_{kl}{}^{KL}
  \right)\big|_{[IJKL]^+} =&\, \frac{2}{3} \delta^{[i}{}_{[k} \left(
    u^{j]m}{}_{IJ} v_{l]m\,KL} - v^{j]m\,IJ} u_{l]m}{}^{KL}
  \right)\Big|_{[IJKL]^+} \\
  \label{eq:u-v-equationsB}
  &- \frac{1}{12} \delta^{ij}_{kl} \left( u^{mn}{}_{IJ} v_{mn\,KL} -
    v^{mn\,IJ} u_{mn}{}^{KL} \right)\big|_{[IJKL]^+},
\end{align}
where $|_{[IJKL]^+}$ represents the projection onto the selfdual part.
This completes the proof of Eq.~(\ref{eq:C-delta}). In order to keep
the formulas short, I do not insert the factorization of the
$\mathcal{C}_{pq}{}^{ijkl}$ tensor into the uplift Ans\"atze for the
warp factor and the three-form. However, one should always keep in
mind that these expressions can still be simplified by
Eq.~(\ref{eq:C-delta}).

I must emphasize that the antisymmetry of the three-form potential
$A_{mnp}$ is not apparent from the new Ansatz in
Eq.~(\ref{eq:A-down}). This may be a hint that it still can be
simplified using the E$_{7(7)}$ properties of the $u_{ij}{}^{IJ}$ and
$v_{ij\,IJ}$ tensors. One should check such a simplification in future
work. Note that the recent three-form Ansatz in \cite{Varela:2015ywx}
is given in a coordinate-free form, hence its components are fully
antisymmetric by definition.

In Section \ref{sec:testing-new-uplift}, I will test the new metric
Ansatz for the G$_2$ invariant solution of 11-dimensional
supergravity. Note that the Ans\"atze for the warp factor and the flux
originate from the old formulas for $\Delta^{-1}g^{mn}$ and
$A_{mn}{}^p$ using the new metric Ansatz. Since these old expressions
were already tested for a G$_2$ invariant solution
\cite{Godazgar:2013nma}, I do not re-check
Eqs.~(\ref{eq:warp-Ansatz},\ref{eq:A-down}) explicitly. For a
consistent test, it will be sufficient to compute the metric by
Eq.~(\ref{eq:metric}) and compare it with the existing expression in
\cite{Godazgar:2013nma}\footnote{Within a G$_2$ invariant solution, an
  expression for the metric has been found by inverting
  $\Delta^{-1}g^{mn}$ `by hand'.}.

\section{A new non-linear Ansatz for the four-form field-strength}
\label{sec:non-linear-ansatz}

In this section, I present a new non-linear Ansatz for the four-form
field-strength
\begin{equation}
  \label{eq:F-definition}
  F_{mnpq} = 4! \, \mathring D_{[m} A_{npq]}.
\end{equation}
So far, the internal three-form potential was only known in particular
cases and it was yet very complicated to compute the derivative of an
explicit expression for $A_{mnp}$. However, I found a new
\textit{general} uplift Ansatz for $A_{mnp}$ in the previous section.
In particular, at the level of 11-dimensional vielbein components
\textbf{(}Eqs.~(\ref{eq:Ansatz-mixed-3form},\ref{eq:pre-metric})\textbf{)},
one finds
\begin{equation}
  \label{eq:A3(V)}
  A_{mnp} = \frac{16\sqrt{2}}{9} \Delta^3 \, \mathcal{V}_{mn}{}^{AB}
  \mathcal{V}_{pq}{}^{[CD} \mathcal{V}^{q\,EF]} \mathcal{V}^r{}_{AB}
  \mathcal{V}_{rs\,CD} \mathcal{V}^s{}_{EF}.
\end{equation}
With a look at Eqs.~(\ref{eq:V11d1},\ref{eq:V11d2}) and using
Eq.~(\ref{eq:Gamma-prop1}), one has
\begin{equation}
  \label{eq:V-prop1}
  \mathcal{V}_{pq}{}^{[CD} \mathcal{V}^{q\,EF]} = \frac{1}{2} \left(
    \mathcal{V}_{pq}{}^{CD} \mathcal{V}^{q\,EF} +
    \mathcal{V}_{pq}{}^{EF} \mathcal{V}^{q\,CD} \right).
\end{equation}
Furthermore, since all SU(8) indices in Eq.~(\ref{eq:A3(V)}) are fully
contracted, I can replace the 11-dimensional vielbeine by the four
dimensional expressions in
Eqs.~(\ref{eq:4d-vielbeinB}-\ref{eq:4d-vielbeinE}). This finally
yields a general expression for the four-form field-strength, i.e.
\begin{equation}
  \label{eq:F-4d}
  F_{mnpq} = \frac{64\sqrt{2}}{3} \, \mathring D_{[m} \left( \Delta^3 \,
    \mathcal{V}_{np}{}^{i_1i_2} \left( \mathcal{V}_{q]r}{}^{i_3i_4}
      \mathcal{V}^{r\,i_5i_6} + \mathcal{V}_{q]r}{}^{i_5i_6}
      \mathcal{V}^{r\,i_3i_4} \right) \mathcal{V}^s{}_{i_1i_2}
    \mathcal{V}_{st\,i_3i_4} \mathcal{V}^t{}_{i_5i_6}\right). 
\end{equation}

One can now evaluate the derivative in general. First, one has
\begin{equation}
  \label{eq:DlogD}
  \mathring D_m \Delta^3 = 3 \Delta^3 \mathring D_m \log \Delta,
\end{equation}
hence, one term in $F_{mnpq}$ will be proportional to $A_{[mnp}
\mathring D_{q]} \log \Delta$. Secondly, the covariant background
derivative $\mathring D_{m}$ only acts on the $y$-dependent fields in
the vielbein components: the Killing forms and the dual volume
potential $\mathring \zeta^m$. \textit{It does not act on the scalars
  $u_{ij}{}^{IJ}$ and $v_{ij\,IJ}$.} In general,
\begin{align}
  \label{eq:DVa}
  \mathring D_m \mathcal{V}^n{}_{ij} =& \, m_7 \, \mathring
  g_{mp} \left( 2 \mathring \zeta^{[n}
    \mathcal{V}^{p]}{}_{ij}  -\mathcal{V}^{np}{}_{ij} 
  \right),\\
  \mathring D_m \mathcal{V}_{np\,ij} =& \, 2m_7\,
  \mathring g_{m[n} \left( - \mathcal{V}_{p]\,ij} + \mathring \zeta^q
    \, \mathcal{V}_{p]q\,ij}
  \right),\\
  \mathring D_m \mathcal{V}^{np}{}_{ij} =& - 2 m_7
  \left( \delta_m{}^{[n} + \mathring \zeta_m \mathring \zeta^{[n} -
    \mathring D_m \mathring \zeta^{[n}
  \right) \mathcal{V}^{p]}{}_{ij} - 2 m_7\,
  \mathring g_{mq} \mathring \zeta^{[n} \mathcal{V}^{p]q}{}_{ij},\\
  \label{eq:DVd}
  \mathring D_m \mathcal{V}_{n\,ij} =& \, m_7 \left(
    \mathring \zeta_m \delta_n{}^p - \mathring g_{mn} \mathring
    \zeta^p \right) \mathcal{V}_{p\,ij} - m_7 \left( \delta_m{}^{p} +
    \mathring \zeta_m \mathring \zeta^{p} - \mathring D_m \mathring
    \zeta^{p} \right) \mathcal{V}_{np\,ij}.
\end{align}

Putting all this together, the resulting intermediate expression for
$F_{mnpq}$ becomes rather long and I do not display it here. However,
it should be clear that it contains the tensors $\mathring g_{mn}$,
$\mathring \zeta^m$ as well as all four-dimensional vielbeine
$\mathcal{V}^\mathcal{M}{}_{ij}$. The SU(8) indices $ij\ldots$ are
fully contracted in pairs. I can therefore replace the
$\mathcal{V}^\mathcal{M}{}_{ij}$'s by the 11-dimensional vielbein
components $\mathcal{V}^\mathcal{M}{}_{AB}$. The final step is to use
Eqs.~(\ref{eq:V11d1}-\ref{eq:V11d4}), which introduces the
11-dimensional fields (e.g. $A_{mnp}$ and $A_{m_1\cdots m_6}$) as well
as $\Gamma$-matrices. Using Eqs.~(\ref{eq:GammaContract}) for the
traces of products of $\Gamma$-matrices, I finally obtain
\begin{align}
  \label{eq:Pre-F(A)}
  F_{mnpq} =& - 72 A_{[mnp} \mathring D_{q]} \log \Delta +
    \frac{24}{\sqrt{2}} m_7 \, A_{[mnp} \mathring g_{q]r} \left(
      \Delta A^r + 3\sqrt{2} \mathring \zeta^r \right) \nonumber\\
    &+ \Big[4 m_7 \mathring g_{m r_1}
    \mathring \eta^{r_1\cdots r_7} \left( g_{n r_2} g_{p r_3} g_{q
        r_4} - 18 A_{np r_2} A_{q r_3 r_4} \right) A_{r_5 r_6 r_7}
  \Big]\Big|_{[mnpq]},
\end{align}
where $|_{[mnpq]}$ denotes antisymmetrized indices $mnpq$. One
eliminates the second term by Eq.~(\ref{eq:A1}),
\begin{align}
  \label{eq:F(A)}
  F_{mnpq} = - 18 A_{[mnp} \mathring D_{q]} \log \Delta + \left[ 4 m_7
    \mathring g_{m r_1} \mathring \eta^{r_1\cdots r_7} \Big( g_{n
        r_2} g_{p r_3} g_{q r_4} - 18 A_{np r_2} A_{q r_3 r_4} \right)
    A_{r_5 r_6 r_7} \Big]\Big|_{[mnpq]}.
\end{align}

For some readers, this expression is already in a desired form.
However, one can further simplify this expression. First, the term
proportional to $\mathring \eta^{r_1\cdots r_7} A_{qr_3r_4}
A_{r_5r_6r_7}$ can be replaced using Eq.~(\ref{eq:A1-ansatz}).
Together with Eq.~(\ref{eq:A1}), this cancels the term proportional to
$\mathring D_m \log \Delta$. Finally, one turns the tensor density
$\mathring \eta^{r_1\cdots r_7}$ into the tensor $\epsilon^{r_1\cdots
  r_7}$ \textbf{(}Eq.~(\ref{eq:tensor-density})\textbf{)} and obtains
\begin{equation}
  \label{eq:new-F-Ansatz}
  F_{mnpq} = m_7 \Delta \mathring g_{s[m}
  \Big(
    4 
    \epsilon_{npq]r_1r_2r_3}{}^s A^{r_1r_2r_3} - 3
    g_{n|t|} A_{pq]r} K^{rs\,IJ} 
    K^{t\,KL}
    \left(
      u_{ij}{}^{IJ} + v_{ij\,IJ}
    \right)
    \left(
      u^{ij}{}_{KL} + v^{ij\,KL}
    \right)
  \Big).
\end{equation}
This formula appears to be more feasible for practical tests than
previous expressions \cite{Godazgar:2015qia,Varela:2015ywx}.

It is not difficult to raise all indices with the inverse metric
$g^{mn}$. Therefore, one must keep in mind that the indices of the
Killing forms and $\mathring g_{mn}$ are raised with the background
metric. All other tensors in Eq.~(\ref{eq:new-F-Ansatz}) are
covariant, hence
\begin{equation}
  \label{eq:new-F-raised}
  F^{mnpq} = m_7 \Delta \mathring g_{st} g^{t[m}\Big( 4
  \epsilon^{npq]r_1r_2r_3s} A_{r_1r_2r_3} - 3 A^{np}{}_r K^{q]\,IJ}
  K^{rs\,KL}
  \left(
    u^{ij}{}_{IJ} + v^{ij\,IJ}
  \right)
  \left(
    u_{ij}{}^{KL} + v_{ij\,KL}
  \right) \Big).
\end{equation}
Note the power of the last step: Until now, the field-strength with
upper indices has always been found by raising each lower index of
$F_{mnpq}$ with the explicit expression for the inverse metric
$g^{mn}$. This was one of the hardest tasks in verifying the
SO(3)$\times$SO(3) invariant solution of 11-dimensional supergravity.
With the new Ansatz above, it is much simpler to find $F^{mnpq}$. For
maximally symmetric spacetimes, these results may also be used to
calculate the Ricci tensor using the Einstein equations.


In the next section, I will test the new Ansatz for the four-form
field-strength for the G$_2$ invariant solution of 11-dimensional
supergravity.

\section{Testing the new uplift Ans\"atze}
\label{sec:testing-new-uplift}

This section presents an essential part of this work: I test the new
non-linear Ans\"atze for the metric $g_{mn}$ and the four-form
field-strength $F_{mnpq}$ within a G$_2$ invariant solution of
11-dimensional supergravity. In such a setup, the Ans\"atze for the
inverse metric $\Delta^{-1}g^{mn}$
\textbf{(}Eq.~(\ref{eq:inverse-metric})\textbf{)} and the three-form
with mixed index structure
\textbf{(}Eq.~(\ref{eq:mixed-3-form})\textbf{)} were already checked
successfully \cite{Godazgar:2013nma}. The same reference computes the
warp factor by taking the determinant of the expression for
$\Delta^{-1}g^{mn}$ and the metric $g_{mn}$ by inverting $g^{mn}$.
Finally, it calculates the full internal three-form potential
$A_{mnp}$ by lowering the third index with the explicit expression for
$g_{mn}$. It should be clear that a successful test for the metric
Ansatz in Eq.~(\ref{eq:metric}) \textit{includes} the tests of the
Ans\"atze for the warp factor and the three-form potential in
Eqs.~(\ref{eq:warp-Ansatz},\ref{eq:A-down}), since these result from
combining the old known Ans\"atze with the new metric Ansatz.

Here, I compute the metric $\Delta^{-2}g_{mn}$ by
Eqs.~(\ref{eq:pre-g},\ref{eq:pre-V}), which is equivalent to use
Eq.~(\ref{eq:metric}). I follow the strategy of
\cite{Godazgar:2013nma}: One first brings the E$_{7(7)}$-matrix that
encodes the four-dimensional scalars
\begin{equation}
  \label{eq:V1}
  \mathcal{V} =
  \begin{pmatrix}
    u_{ij}{}^{IJ} & v_{ij\,IJ} \\ v^{ij\,IJ} & u^{ij}{}_{IJ}
  \end{pmatrix}
\end{equation}
into unitary gauge,
\begin{equation}
  \label{eq:V2}
  \mathcal{V} = \exp
  \begin{pmatrix}
    0 & \phi_{IJKL} \\ \phi^{IJKL} & 0
  \end{pmatrix},
\end{equation}
where $\phi_{IJKL}$ denotes the scalar vacuum expectation value. In
this gauge, there is no distinction between SU(8) indices $ij\ldots$
and SL(8,$\mathbb{R}$) indices $IJ\ldots$ This allows us to write the
scalar fields $u_{ij}{}^{IJ}$ and $v_{ij\,IJ}$ in terms of the vacuum
expectation value $\phi_{IJKL}$. For a G$_2$ invariant configuration,
the latter takes the general form,
\begin{equation}
  \label{eq:phi}
  \phi_{IJKL}(x) = \frac{\lambda(x)}{2}
  \left(
    C_+^{IJKL} \cos \alpha(x) + i C_-^{IJKL} \sin \alpha(x)
  \right),
\end{equation}
where $C_+^{IJKL}$ is selfdual and $C_-^{IJKL}$ is anti-selfdual. The
above expression also defines a scalar field $\lambda(x)$ and a
rotation angle $\alpha(x)$. Using the explicit form of the vacuum
expectation value in Eq.~(\ref{eq:phi}), one finds the
four-dimensional scalars $u_{ij}{}^{IJ}$ and $v_{ij\,IJ}$ in terms of
the G$_2$ invariants $C_\pm^{IJKL}$, i.e.
\begin{gather}
  \label{eq:u-->CD}
  u_{IJ}{}^{KL} = p^3 \delta^{IJ}_{KL} + \frac{1}{2} p q^2
  \cos^2\alpha C_+^{IJKL} - \frac{1}{2} p q^2 \sin^2\alpha C_-^{IJKL}
  - \frac{i}{8} p q^2 \sin 2\alpha D_-^{IJKL},\\
  \label{eq:v-->CD}
  v_{IJKL} = q^3 (\cos^3\alpha - i \sin^3\alpha ) \delta^{IJ}_{KL} +
  \frac{1}{2} p^2 q \cos \alpha C_+^{IJKL} + \frac{i}{2} p^2 q \sin
  \alpha C_-^{IJKL} - \frac{1}{8} q^3 \sin 2\alpha ( \sin \alpha - i
  \cos \alpha ) D_+^{IJKL},
\end{gather}
where $p = \cosh \lambda$ and $q = \sinh \lambda$. The tensors
$D_\pm^{IJKL}$ are defined as
\begin{equation}
  \label{eq:D}
  D_{\pm}^{IJKL} = \frac{1}{2}
  \left(
    C_+^{IJMN} C_-^{MNKL} \pm C_-^{IJMN} C_+^{MNKL}
  \right).
\end{equation}

One now expands the $C_\pm^{IJKL}$ tensors into the (anti-)selfdual
bases provided by the Killing forms defined in
Eq.~(\ref{eq:K-selfdual-basis},\ref{eq:K-antiselfdual-basis}),
\begin{gather}
  \label{eq:C+-->K}
  C_+^{IJKL} = \frac{\xi}{6} K_m{}^{[IJ} K^{m\,KL]} + \frac{1}{12}
  \xi^m K_{mn}{}^{[IJ} K^{n\,KL]} - \frac{3}{2} K_m{}^{[IJ}
  K_n{}^{KL]} \xi^{mn},\\
  \label{eq:C--->K}
  C_-^{IJKL} = \frac{1}{2} S^{mnp} K_{mn}{}^{[IJ} K_p{}^{KL]}.
\end{gather}
The occurring components $\xi$, $\xi^m$, $\xi^{mn}$ and $S^{mnp}$ are
SO(7) tensors\footnote{In \cite{Godazgar:2013nma}, $S^{mnp}$ was
  denoted by $\mathring S^{mnp}$.} on the round $S^7$, hence, its
indices are raised and lowered with the background metric $\mathring
g_{mn}$. Note that $S^{mnp}$ is totally antisymmetric by construction.
Furthermore, one finds the useful relations \cite{Godazgar:2013nma}
\begin{gather}
  \label{eq:XiIdentities}
  \xi^{mn} \mathring g_{mn} = \xi,\qquad \xi_m \xi_n = (9 - \xi^2)
  \mathring g_{mn} - 6(3 - \xi) \xi_{mn},
  \qquad \xi_m \xi^m = (21 + \xi)(3 - \xi) ,\\
  \label{eq:SIdentities}
  S^{mnr} S_{pqr} = 2 \delta^{mn}_{pq} + \frac{1}{6} \mathring
  \eta^{mn}{}_{pqrst} S^{rst}, \ S^{[mnp} S^{q]rs} = \frac{1}{4}
  \mathring \eta^{mnpq [r}{}_{tu} S^{s]tu}, \ S^{m[np} S^{qr]s} =
  \frac{1}{6} \mathring \eta^{npqr(m}{}_{tu} S^{s)tu}.
\end{gather}

From the decomposition of the $C_\pm^{IJKL}$ tensors in
Eqs.~(\ref{eq:C+-->K},\ref{eq:C--->K}), one finds the useful
contractions
\begin{align}
  C_+^{IJKL} K_m{}^{KL} =& - 2 \xi_{mn} K^{n\,IJ} - \frac{1}{3} \xi^n
  K_{mn}{}^{IJ},\\
  C_+^{IJKL} K_{mn}{}^{KL} =&\, \frac{2}{3} \xi_{[m} K_{n]}{}^{IJ} +
  \left( \frac{2}{3} \xi \delta^{pq}_{mn} - 4 \delta_{[m}{}^p
    \xi_{n]}{}^q \right) K_{pq}{}^{IJ},\\ 
  C_-^{IJKL} K_m{}^{KL} =& \,S_{mnp} K^{np\,IJ},\\
  C_-^{IJKL} K_{mn}{}^{KL} =& \, 2 S_{mnp} K^{p\,IJ} - \frac{1}{6}
  \mathring \eta_{mnp_1\cdots p_5} S^{p_1p_2p_3} K^{p_4p_5\,IJ},
\end{align}
as well as for the $D_\pm^{IJKL}$ tensors,
\begin{align}
  D_+^{IJKL} K_m{}^{KL} =& \left( \frac{\xi}{3} S_{mnp} - \xi_m{}^q
    S_{npq} - 2 S_{mnq} \xi_p{}^q + \frac{1}{36} \mathring
    \eta_{mnpqrst} \xi^q S^{rst} \right) K^{np\,IJ},\\
  D_+^{IJKL} K_{mn}{}^{KL} =& \left( \frac{2}{3} \xi S_{mnp} - 4
    \xi_{[m}{}^q S_{n]pq} - 2 S_{mnq} \xi_p{}^q + \frac{1}{18}
    \mathring \eta_{mnpqrst}
    \xi^q S^{rst} \right) K^{p\,IJ} + \nonumber\\
  & + \frac{1}{3} \left( \xi_{[m} S_{n]pq} - S_{mnp} \xi_q -
    \frac{\xi}{3} \mathring \eta_{mnpqrst} S^{rst} + \mathring
    \eta_{mnprstu} \xi_q{}^r S^{stu} + \mathring \eta_{[m|pqrstu}
    \xi_{n]}{}^{r} S^{stu} \right) K^{pq\,IJ}\\
  D_-^{IJKL} K_m{}^{KL} =& \, \frac{2}{3} S_{mnp} \xi^n K^{p\,IJ} +
  \left( \frac{\xi}{3} S_{mnp} + \xi_m{}^q S_{npq} - 2 S_{mnq}
    \xi_{p}{}^{q} - \frac{1}{36} \mathring \eta_{mnpqrst} \xi^q
    S^{rst}
  \right) K^{np\,IJ},\\
  D_-^{IJKL} K_{mn}{}^{KL} =& \left( -\frac{2}{3} \xi S_{mnp} + 4
    \xi_{[m}{}^q S_{n]pq} - 2 S_{mnq} \xi_{p}{}^{q} + \frac{1}{18}
    \mathring \eta_{mnpqrst} \xi^q S^{rst}
  \right) K^{p\,IJ} + \nonumber\\
  & + \frac{1}{3} \left( - \xi_{[m} S_{n]pq} - S_{mnp} \xi_q +
    \mathring \eta_{mnprstu} \xi_{q}{}^{r} S^{stu} - \mathring
    \eta_{[m|pqrstu} \xi_{n]}{}^{r} S^{stu} \right) K^{pq\,IJ}.
\end{align}

Now, I write the metric $g_{mn}$ in terms of the components $\xi$,
$\xi^m$, $\xi^{mn}$ and $S^{mnp}$ defined above. Therefore, one first
computes $\mathcal{V}_{mp\,[ij}\mathcal{V}^{p}{}_{kl]}$ \textbf{(}or
better: $\mathcal{V}_{mp\,[IJ}\mathcal{V}^{p}{}_{KL]}$\textbf{)} using
Eq.~(\ref{eq:pre-V}) and expands the scalar fields $u_{IJ}{}^{KL}$ and
$v_{IJ\,KL}$ in terms of the $C_\pm^{IJKL}$ and $D_\pm^{IJKL}$ tensors
\textbf{(}Eqs.~(\ref{eq:u-->CD},\ref{eq:v-->CD})\textbf{)}. Secondly,
one uses the contractions above together with
Eqs.~(\ref{eq:K-prop3},\ref{eq:K-prop4}) in Appendix
\ref{sec:kill-spin-kill} to bring
$\mathcal{V}_{mp\,[IJ}\mathcal{V}^{p}{}_{KL]}$ into the basis provided
by Eqs.~(\ref{eq:K-selfdual-basis},\ref{eq:K-antiselfdual-basis}),
\begin{equation}
  \label{eq:4}
  \mathcal{V}_{mp\,[IJ}\mathcal{V}^{p}{}_{KL]} = a_m K_{n}{}^{[IJ}
  K^{n\,KL]} + b_m{}^n K_{np}{}^{[IJ} K^{p\,KL]} + c_{m}{}^{np}
  K_n{}^{[IJ} K_p{}^{KL]} + d_m{}^{npq} K_{[np}{}^{[IJ}
  K_{q]}{}^{KL]}. 
\end{equation}
The respective coefficients $a_m$, $b_m{}^n$, $c_m{}^{np}$ and
$d_m{}^{npq}$ are rather long expressions and I do not display them
here. However, it should be clear that they only depend on the SO(7)
tensors $\xi$, $\xi^m$, $\xi^{mn}$ and $S^{mnp}$. Finally, one
computes the metric via Eq.~(\ref{eq:pre-g}). For the contractions of
the indices $IJKL$, one uses
Eqs.~(\ref{eq:K-prop5}-\ref{eq:K-Gamma-0}) and for the contractions of
the SO(7) indices, one uses the identities in
Eqs.~(\ref{eq:XiIdentities},\ref{eq:SIdentities}). This finally
results in
\begin{equation}
  \label{eq:metric-G2}
  \Delta^{-2} g_{mn} = b_0 \big[
  \left(
    b_0 + 3 cvs
  \right) \mathring g_{mn} + cvs \, \xi_{mn} \big],
\end{equation}
where I made the following definitions:
\begin{gather}
  \label{eq:defs}
  c = \cosh 2\lambda ,\qquad s = \sinh 2\lambda, \qquad v = \cos
  \alpha, \qquad b_0 = c^2 + v^2s^2 - \frac{9+\xi}{6}\, cvs.
\end{gather}

The test for the inverse metric Ansatz
\textbf{(}Eq.~(\ref{eq:inverse-metric})\textbf{)} was already
performed in \cite{Godazgar:2013nma}. The corresponding final
expression is
\begin{equation}
  \label{eq:inverse-metric-G2}
  \Delta^{-1} g^{mn} = \left(c^3 + v^3s^3\right) \mathring g^{mn} -
  cvs (c + vs) \xi^{mn}.
\end{equation}
Combining the explicit expressions for the metric and its inverse in
Eqs.~(\ref{eq:metric-G2},\ref{eq:inverse-metric-G2}) and using the
identities in Eqs.~(\ref{eq:XiIdentities},\ref{eq:SIdentities}), one
finds that
\begin{equation}
  \label{eq:warp-G2-pre}
  \Delta^{-2} g_{mp} \Delta^{-1} g^{pn} = b_0^2\,(c + vs)^3 \delta_m^n.
\end{equation}
This is exactly the combination of the metric and its inverse that
defines the warp factor in Eq.~(\ref{eq:warp-Ansatz}), hence
\begin{equation}
  \label{eq:warp-G2}
  \Delta^{-3} = b_0^2\,(c + vs)^3.
\end{equation}
The explicit expressions for the metric and the warp factor in
Eqs.~(\ref{eq:metric-G2},\ref{eq:warp-G2}) reproduce the results of
\cite{Godazgar:2013nma}. The reader may also check that the
determinant of the metric in Eq.~(\ref{eq:metric-G2}) indeed,
reproduces Eq.~(\ref{eq:warp-G2}). The test is hence, successful.

For the remaining test of the field-strength Ansatz in
Eq.~(\ref{eq:new-F-Ansatz}), I use the explicit expression for
$A_{mnp}$ that was found in \cite{Godazgar:2013nma},
\begin{equation}
  \label{eq:A3}
  A_{mnp} = \frac{\sqrt{2}\, \tan \alpha}{72b_0} \frac{vs}{c+vs}
  \left(
    9vs(c-vs) \xi_{[m}{}^q S_{np]q} + \frac{1}{12} vs (c+vs) \mathring
    \eta_{mnpqrst} \xi^q S^{rst} + (2c - vs) (3c - \xi vs) S_{mnp}
  \right).
\end{equation}
Note that this expression is slightly simplified using the identities
in Eqs.~(\ref{eq:XiIdentities},\ref{eq:SIdentities}). Furthermore, the
formula for $A_{mnp}$ above differs from the expression given in
\cite{Godazgar:2013nma} by a factor of $1/6$, which is due to my
conventions. However, the definition of the field-strength in
Eq.~(\ref{eq:F-definition}) differs from the corresponding definition
in \cite{Godazgar:2013nma} by a factor of 6. Hence, the new Ansatz for
$F_{mnpq}$ in Eq.~(\ref{eq:new-F-Ansatz}) should give the same
expression as already computed in \cite{Godazgar:2013nma} by
calculating the derivative of Eq.~(\ref{eq:A3}) directly.

A convenient way to use the new Ansatz is
\begin{align}
  \label{eq:F-convenient}
  F_{mnpq} =& \Big[4 m_7 \Delta^6 \mathring g_{mr_1}
  \left(\Delta^{-2}g_{nr_2}\right) \left(\Delta^{-2}g_{pr_3}\right)
  \left(\Delta^{-2}g_{qr_4}\right) \mathring \eta^{r_1 \cdots r_7}
  A_{r_5 r_6 r_7} - \nonumber\\
  & \qquad\qquad - 3 m_7 \Delta^3 \left( \Delta^{-2}g_{nt} \right) A_{pqr}
  K^{rs\,IJ} K^{t\,KL} \left( u_{ij}{}^{IJ} + v_{ij\,IJ} \right)
  \left( u^{ij}{}_{KL} + v^{ij\,KL} \right) \Big] \Big|_{[mnpq]},
\end{align}
such that one may use
Eqs.~(\ref{eq:metric-G2},\ref{eq:warp-G2},\ref{eq:A3}) directly. For
the term involving the Killing forms and the four-dimensional scalars,
I follow the same strategy as described earlier in this section. I
find
\begin{align}
  \label{eq:H3}
  K^{mn\,IJ} K^{p\,KL} \left( u_{ij}{}^{IJ} + v_{ij\,IJ} \right)&
  \left( u^{ij}{}_{KL} + v^{ij\,KL} \right) = \frac{8}{3} cvs\,
  (c+vs)\, \xi^{[m}\mathring g^{n]p}
  +\nonumber\\
  & + s^2 \sin^2\alpha \left[ 12 \, vs \, \xi^{[m}{}_q S^{np]q} -
    \frac{1}{9} vs \, \mathring \eta^{mnpqrst} \xi_q S_{rst} -
    \left(8c + \frac{4}{3} \xi vs \right) S^{mnp} \right].
\end{align}
Putting all together and using
Eqs.~(\ref{eq:XiIdentities},\ref{eq:SIdentities}) finally results in
\begin{align}
  \label{eq:F-G2}
  F_{mnpq} = \frac{\sqrt{2} v^2 s^2 \tan \alpha}{3b_0} \,m_7
  \bigg[
    \frac{c-vs}{vs} \, \mathring \eta_{mnpqrst} S^{rst} +&
    \left(
      \frac{2c-vs}{c+vs} + \frac{c^2 - v^2s^2}{b_0}
    \right) \xi_{[m} S_{npq]} + \nonumber\\
    +& \frac{1}{6(3-\xi)}
    \left(
      \frac{2c-vs}{c+vs} - \frac{(c - vs)^2}{b_0}
    \right) \xi_{[m} \mathring \eta_{npq]rstu} \xi^r S^{stu}
  \bigg],
\end{align}
which matches exactly the expression found in \cite{Godazgar:2013nma}.
The test is hence, successful.

\section{Conclusion}
\label{sec:conclusion}

In this paper, I derive a new non-linear metric Ansatz for the uplift
of $N=8$ supergravity to 11-dimensional supergravity. An uplift Ansatz
for the inverse metric, scaled with the warp factor,
$\Delta^{-1}g^{mn}$, has already been known for a long time
\cite{deWit:1984nz}. However, inverting this expression in order to
find $\Delta g_{mn}$ was only possible in certain cases, for example
when the theory is G$_2$, SO(3)$\times$SO(3) or
SU(3)$\times$U(1)$\times$U(1) invariant
\cite{Godazgar:2013nma,Godazgar:2014eza,Pilch:2015vha,Pilch:2015dwa}.
Also the warp factor $\Delta$ could only be extracted by taking the
determinant in such particular cases. Following the strategy of
\cite{Godazgar:2013pfa}, I present a new \textit{general} uplift
Ansatz for $\Delta^{-2}g_{mn}$ in terms of the four-dimensional scalar
fields and the Killing forms on the background
\textbf{(}Eqs.~(\ref{eq:pre-g}-\ref{eq:metric})\textbf{)}. Note that
this Ansatz is similar to a recent coordinate-free expression
\cite{Varela:2015ywx}. However, the formula presented here seems to be
more feasible for practical tests: I tested the new metric Ansatz
within a G$_2$ invariant solution of 11-dimensional supergravity in
Section \ref{sec:testing-new-uplift}.

Similarly to \cite{Varela:2015ywx}, the new formula can further be
used in order to find non-linear uplift Ans\"atze for the warp factor
and the full internal three-form potential in general. For the warp
factor, I combine the old Ansatz for $\Delta^{-1}g^{mn}$ with the new
one for $\Delta^{-2}g_{mn}$, which gives a new Ansatz for
$\Delta^{-3}$ \textbf{(}Eq.~(\ref{eq:warp-Ansatz})\textbf{)}.
Furthermore, I derive a general Ansatz for the full internal
three-form potential $A_{mnp}$
\textbf{(}Eq.~(\ref{eq:A-down})\textbf{)} by combining the old flux
Ansatz for $A_{mn}{}^p$ \cite{Godazgar:2013pfa} with the new metric
Ansatz. However, this new formula does not reveal the total
antisymmetry of the three-form. This may be a hint that one can
further simplify the expression for $A_{mnp}$ using some E$_{7(7)}$
identities for the four-dimensional scalar fields. I hope that I can
provide such a simplification in future work.

In a second part of this paper, I derive a new general non-linear
uplift Ansatz for the four-form field-strength $F_{mnpq}$ within the
considered uplift of $N=8$ supergravity to 11-dimensional
supergravity. So far, the simplest way to derive $F_{mnpq}$ was to
compute the derivative of the three-form potential. However, this
required an explicit expression for the flux, which is only given in
particular cases, e.g. the G$_2$, SO(3)$\times$SO(3) or
SU(3)$\times$U(1)$\times$U(1) invariant solutions of 11-dimensional
supergravity. With the new Ansatz for the field-strength
\textbf{(}Eq.~(\ref{eq:new-F-Ansatz})\textbf{)}, there is no need to
compute derivatives anymore. It is given in terms of the metric, the
flux as well as the four-dimensional scalars and background Killing
forms. The formula holds in general and also passes a very non-trivial
test for a G$_2$ invariant solution of 11-dimensional supergravity.

The new Ansatz for the field-strength also provides a simple
expression for $F^{mnpq}$
\textbf{(}Eq.~(\ref{eq:new-F-raised})\textbf{)} in terms of the
inverse metric, the flux as well as the four-dimensional scalars and
background Killing forms. This new formula makes it redundant to raise
each index of $F_{mnpq}$ with the explicit expression for the inverse
metric, $g^{mn}$, which was so far, the only way to derive $F^{mnpq}$.
The new direct Ansatz for $F^{mnpq}$ is also much more effective than
this old method --- in order to verify the SO(3)$\times$SO(3)
invariant solution of 11-dimensional supergravity, the index-raising
of $F_{mnpq}$ was one of the hardest tasks \cite{Godazgar:2015qia}.

In future, one may also find new Ans\"atze for the Christoffel
connections in 11-dimensional supergravity in terms of the
four-dimensional scalars and background Killing forms. Since they are
given by the first derivative of the metric, one could find new simple
expressions in full analogy to the derivation of the field-strength
Ansatz. Similarly, one could derive a non-linear Ansatz for the
Riemann tensor.

In this paper, all Ans\"atze are derived within the S$^7$ reduction of
11-dimensional supergravity. This leads to the compact gauging SO(8).
However, the methods provided here should also apply in general for
other truncations. As a first example, one may extend the theory to
the non-compact CSO($p,q,r$) gaugings
\cite{Hull:1988jw,Baron:2014bya}. In this case, the $IJ$ indices of
the Killing forms are raised and lowered with the CSO($p,q,r$)-metric
$\eta_{IJ}$ instead of the SO(8) metric $\delta_{IJ}$. This effects
the definition of the matrix $\mathcal{R}^\mathcal{M}{}_\mathcal{N}$
in Eqs.~(\ref{eq:R-explicitA}-\ref{eq:R-explicitD}) and hence, the
$\mathcal{A}_{m\,ijkl}$ and $\mathcal{B}_{m\,ijkl}$ tensors in
Eqs.~(\ref{eq:A},\ref{eq:B}). Thus, the new Ans\"atze for the metric,
the three-form and warp-factor will be slightly modified. However, the
new Ansatz for the four-form field-strength will change more
dramatically: Eqs.~(\ref{eq:DVa}-\ref{eq:DVd}) do not hold if the $IJ$
indices of the Killing forms are raised and lowered with the full
CSO($p,q,r$) metric. Since the new Ansatz for $F_{mnpq}$ depends on
those identities, it will take much more effort to derive an adapted
Ansatz for the four-form within the non-compact gaugings. Finally, the
presented methods may also be used for the reduction from type IIB
supergravity to five dimensions
\cite{Lee:2014mla,Ciceri:2014wya,Baguet:2015sma}.

\section*{Acknowledgments}
\label{sec:acknowledgments}

I thank Hermann Nicolai, Axel Kleinschmidt as well as Hadi and Mahdi
Godazgar for useful discussions.

\appendix

\section{Gamma matrices, Killing spinors, Killing vectors and Killing
  forms of the $S^7$}
\label{sec:kill-spin-kill}

One defines a set of euclidean, antisymmetric and purely imaginary
$8\times8$ $\Gamma$-matrices ($\Gamma^\dagger = \Gamma$). These
generate the euclidean Clifford algebra in seven
dimensions\footnote{In the following, I suppress the SU(8) indices
  that label rows and columns of the matrices, $\Gamma_a = \left(
    \Gamma_a \right)_{AB}$.},
\begin{eqnarray}
  \label{eq:clifford}
  \{\Gamma_a,\Gamma_b \} = 2 \delta_{ab} \mathbb{I}_{8\times8}.
\end{eqnarray}
Let us choose a Majorana representation: The charge conjugation matrix
that defines spinor conjugates or raises and lowers spinor indices is
set to the unit matrix. Thus, the eight Killing spinors of the round
$S^7$ satisfy $\bar \eta^I = (\eta^I)^\dagger$. Furthermore, one may
choose them to be orthonormal,
\begin{eqnarray}
  \label{eq:spinor-orthonormal}
  \bar \eta^I \eta^J = \delta^{IJ}, \quad \eta^I \bar \eta^I =
  \mathbb{I}_{8\times8}.
\end{eqnarray}

The flat $\Gamma$-matrices define two types of `curved'
$\Gamma$-matrices: First, matrices $\mathring \Gamma_m = \mathring
e_m^a \Gamma_a$ are defined on the round seven-sphere, its indices are
raised and lowered with the background metric $\mathring g_{mn}$.
Secondly, matrices $\Gamma_m = e_m^a \Gamma_a$ are defined on the
\textit{deformed} $S^7$ and its indices are raised and lowered with
the full internal metric $g_{mn}$.

The Killing spinors are defined on the background $S^7$ and hence,
satisfy
\begin{eqnarray}
  \label{eq:dirac}
  i \mathring D_m \eta^I = \frac{m_7}{2} \mathring \Gamma_m
  \eta^I, \quad   -i \mathring D_m \bar\eta^I = \frac{m_7}{2} \bar
  \eta^I \mathring \Gamma_m.
\end{eqnarray}
Here, $\mathring D_m$ is the covariant derivative with respect to the
internal background metric $\mathring g_{mn}$ and $m_7$ is the inverse
$S^7$ radius.

The $\Gamma$-matrices can be used to define two sets of $8\times8$
matrices,
\begin{eqnarray}
  \label{eq:general-gammas}
  \mathring \Gamma_{m_1 \ldots m_i} = \mathring \Gamma_{[ m_1} \ldots
  \mathring \Gamma_{m_i]}, \qquad\qquad \Gamma_{m_1 \ldots m_i} =
  \Gamma_{[ m_1} \ldots \Gamma_{m_i]} 
\end{eqnarray}
for $i = 2, \ldots, 7$. For example,
\begin{equation}
  \label{eq:Gamma-example}
  \Gamma_{mnp} = \frac{1}{3!}
  \left(
    \Gamma_{mnp} + \Gamma_{npm} + \Gamma_{pmn} - \Gamma_{mpn} -
    \Gamma_{nmp} - \Gamma_{pnm}
  \right).
\end{equation}
$\Gamma$-matrices with one and two indices are antisymmetric and
$\Gamma$-matrices with three indices are symmetric. The two sets $
\begin{pmatrix}
  \mathbb{I}_{8\times8},& \mathring \Gamma_m,& \mathring \Gamma_{mn},&
  \mathring \Gamma_{mnp}
\end{pmatrix}
$ and $
\begin{pmatrix}
  \mathbb{I}_{8\times8},& \Gamma_m,& \Gamma_{mn},& \Gamma_{mnp}
\end{pmatrix}
$ each contain $1 + 7 + 21 + 35 = 64$ independent matrices. Hence,
they both span the vector space of $8\times8$ matrices. In these
bases,
\begin{gather}
  \label{eq:dual-gammas7}
  \mathring \Gamma_{m_1 \ldots m_7} = -i \mathring \eta_{m_1 \ldots
    m_7} \mathbb{I}_{8\times8}, \qquad\qquad \Gamma_{m_1 \ldots m_7} =
  -i \epsilon_{m_1 \ldots m_7} \mathbb{I}_{8\times8},\\
  \label{eq:dual-gammas6}
  \mathring \Gamma_{m_1 \ldots m_6} = -i \mathring \eta_{m_1 \ldots
    m_7} \mathring \Gamma^{m_7}, \qquad\qquad \Gamma_{m_1 \ldots m_6}
  = -i \epsilon_{m_1 \ldots m_7}  \Gamma^{m_7},\\
  \label{eq:dual-gammas5}
  \mathring \Gamma_{m_1 \ldots m_5} = \frac{i}{2} \mathring \eta_{m_1
    \ldots m_7 } \mathring \Gamma^{m_6 m_7}, \qquad\qquad \Gamma_{m_1
    \ldots m_5} = \frac{i}{2} \epsilon_{m_1 \ldots m_7 }
  \Gamma^{m_6 m_7},\\
  \label{eq:dual-gammas4}
  \mathring \Gamma_{m_1 \ldots m_4} = \frac{i}{3!} \mathring \eta_{m_1
    \ldots m_7 } \mathring \Gamma^{m_5\cdots m_7}, \qquad\qquad
  \Gamma_{m_1 \ldots m_4} = \frac{i}{3!} \epsilon_{m_1 \ldots m_7 }
  \Gamma^{m_5\cdots m_7}.
\end{gather}
Beside the Clifford algebra, the $\Gamma$-matrices satisfy the useful
relations
\begin{gather}
  \label{eq:KContract}
  \mathrm{Tr} \left(\mathring \Gamma^m \mathring \Gamma^n \right) = 8
  \mathring g^{mn},\qquad \mathrm{Tr} \left(\mathring \Gamma^m
    \mathring \Gamma^{np}\right) = 0,\qquad \mathrm{Tr}
  \left(\mathring \Gamma^{mn} \mathring
    \Gamma_{pq} \right) = -16\delta^{mn}_{pq},\\
  \label{eq:GammaContract}
  \mathrm{Tr} \left( \Gamma^m \Gamma^n \right) = 8 g^{mn},\qquad
  \mathrm{Tr} \left( \Gamma^m \Gamma^{np}\right) = 0,\qquad
  \mathrm{Tr} \left( \Gamma^{mn} \Gamma_{pq} \right) =
  -16\delta^{mn}_{pq}.
\end{gather}

The Killing spinors define a set of Killing vectors and their
derivatives,
\begin{align}
  \label{eq:Killings}
  K_m{}^{IJ} =&\, i\bar\eta^I \mathring\Gamma_m \eta^J, \\
  K_{mn}{}^{IJ} =&\, \bar\eta^I \mathring \Gamma_{mn} \eta^J,\\
  K_{m_1\cdots m_5}{}^{IJ} =&\, i \bar\eta^I \mathring \Gamma_{m_1\cdots m_5}
  \eta^J.
\end{align}
Using Eq.~(\ref{eq:dirac}), one verifies that $K_{mn}{}^{IJ}$ is
indeed, proportional to the derivative of $K_m{}^{IJ}$,
\begin{equation}
  \label{eq:Killing-derivatives}
  \mathring D_n K_m{}^{IJ} = m_7 K_{mn}{}^{IJ}, \qquad
  \mathring D_p K_{mn}{}^{IJ} = 2 m_7 \mathring g_{p[m} K_{n]}{}^{IJ}.
\end{equation}
Using Eq.~(\ref{eq:dual-gammas5}), one also finds that $K_{m_1\cdots
  m_5}{}^{IJ}$ is the (seven dimensional) dual to $K_{mn}{}^{IJ}$,
\begin{equation}
  \label{eq:Killing-dual}
  K_{m_1\cdots m_5} = -\frac{1}{2} \mathring \eta_{m_1\cdots m_7}
  K^{m_6 m_7\;IJ}. 
\end{equation}
Note that curved seven dimensional indices of the Killing vectors and
their derivatives are always raised and lowered with the background
metric $\mathring g_{mn}$.

The following bi-linears in the $\Gamma$-matrices represent a basis
for (anti-)selfdual SU(8) tensors on the deformed seven-sphere:
\begin{gather}
  \label{eq:Gamma-selfdual-basis}
  \mathrm{selfdual}: \quad \Gamma_{m\,[AB} \Gamma^m{}_{CD]}, \qquad
  \Gamma_{mn\,[AB} \Gamma^n{}_{CD]},\qquad \Gamma^m{}_{[AB}
  \Gamma^n{}_{CD]}\\ 
  \label{eq:Gamma-antiselfdual-basis}
  \mathrm{anti-selfdual}: \quad \Gamma^{[mn}{}_{[AB}
  \Gamma^{p]}{}_{CD]}.
\end{gather}
On the background, there is the respective basis of (anti-)selfdual
SL(8) tensors in terms of the Killing bi-linears, i.e.
\begin{gather}
  \label{eq:K-selfdual-basis}
  \mathrm{selfdual}: \quad K_m{}^{[IJ} K^{m\,KL]}, \qquad
  K_{mn}{}^{[IJ} K^{n\,KL]},\qquad K_m{}^{[IJ} K_n{}^{KL]}\\
  \label{eq:K-antiselfdual-basis}
  \mathrm{anti-selfdual}: \quad  K_{[mn}{}^{[IJ} K_{p]}{}^{KL]}.
\end{gather}
These bi-linears satisfy further useful relations
\cite{deWit:1986mz,Godazgar:2013nma},
\begin{align}
  \label{eq:Gamma-prop0}
  \Gamma_{m\,[AB} \Gamma^m{}_{CD]} =&\, \Gamma_{m\,AB} \Gamma^m{}_{CD} +
  2 \delta^{AB}_{CD},\\
  \label{eq:Gamma-prop1}
  \Gamma_{mn\,[AB} \Gamma^n{}_{CD]} =&\, \frac{1}{2} \left(
    \Gamma_{mn\,AB} \Gamma^n{}_{CD} + \Gamma_{mn\,CD} \Gamma^n{}_{AB}
  \right),\\
  \label{eq:Gamma-prop3}
  \Gamma^{mn}{}_{[AB} \Gamma^p{}_{CD]} =& -\frac{1}{3} g^{p[m}
  \Gamma^{n]q}{}_{[AB} \Gamma_{q\,CD]} + \Gamma^{[mn}{}_{[AB}
  \Gamma^{p]}{}_{CD]},\\
  \label{eq:Gamma-prop4}
  \Gamma^{mn}{}_{[AB} \Gamma^{pq}{}_{CD]} =& - 2 g^{m[p}
  \Gamma^{q]}{}_{[AB} \Gamma^{n}{}_{CD]} + 2 g^{n[p}
  \Gamma^{q]}{}_{[AB} \Gamma^{m}{}_{CD]} +\nonumber\\
  & + \frac{2}{3} g^{m[p} g^{q]n} \Gamma_{r\,[AB} \Gamma^r{}_{CD]} +
  \Gamma^{[mn}{}_{[AB} \Gamma^{pq]}{}_{CD]}
\end{align}
as well as
\begin{align}
  \label{eq:K-prop0}
  K_m{}^{[IJ} K^{m\,KL]} =&\, K_m{}^{IJ} K^{m\,KL} - 2\delta^{IJ}_{KL},\\
  \label{eq:K-prop1}
  K_{mn}{}^{[IJ} K^n{}^{KL]} =&\, \frac{1}{2} \left( K_{mn}{}^{IJ}
    K^n{}^{KL} + K_{mn}{}^{KL}
    K^n{}^{IJ} \right),\\
  \label{eq:K-prop3}
  K_{mn}{}^{[IJ} K_p{}^{KL]} =& -\frac{1}{3} \mathring g_{p[m}
  K_{n]q}{}^{[IJ} K^{q\,KL]} + K_{[mn}{}^{[IJ} K_{p]}{}^{KL]},\\
  \label{eq:K-prop4}
  K_{mn}{}^{[IJ} K_{pq}{}^{KL]} =&\, 2 \mathring g_{m[p}
  K_{q]}{}^{[IJ} K_{n}{}^{KL]} - 2 \mathring g_{n[p}
  K_{q]}{}^{[IJ} K_{m}{}^{KL]} -\nonumber\\
  & - \frac{2}{3} \mathring g_{m[p} \mathring g_{q]n} K_r{}^{[IJ}
  K^{r\,KL]} + K_{[mn}{}^{[IJ} K_{pq]}{}^{KL]}.
\end{align}
One has furthermore \cite{deWit:1986mz}
\begin{align}
  \label{eq:Gamma-prop2}
  \Gamma_{mn\,AB} \Gamma^n{}_{CD} - \Gamma_{mn\,CD} \Gamma^n{}_{AB} =&
  -4
  \left(
    \delta_{C[A} \Gamma_{m\,B]D} - \delta_{D[A} \Gamma_{m\,B]C}
  \right),\\
  \label{eq:K-prop2}
  K_{mn}{}^{IJ} K^n{}^{KL} - K_{mn}{}^{KL} K^n{}^{IJ} =& -8
  \delta^{[I}{}_{[K} K_m{}^{J]}{}_{L]}.
\end{align}

The bases in
Eqs.~(\ref{eq:Gamma-selfdual-basis}-\ref{eq:K-antiselfdual-basis}) are
in some sense `orthogonal'. Indeed, one has
\begin{align}
  \label{eq:Gamma-prop5}
  \Gamma^m{}_{[AB} \Gamma^n{}_{CD]} \Gamma^p{}_{AB}
  \Gamma^q{}_{CD} =&\, 16 g^{m(p} g^{nq)},\\[2pt]
  \label{eq:Gamma-prop6}
  \Gamma_{mp\,[AB} \Gamma^p{}_{CD]} \Gamma_{nq\,AB}
  \Gamma^q{}_{CD} =&- 192 g_{mn},\\[2pt]
  \label{eq:Gamma-prop7}
  \Gamma^{[mn}{}_{[AB} \Gamma^{p]}{}_{CD]} \Gamma_{[qr\,AB}
  \Gamma_{s]CD} =&- 32 \delta^{mnp}_{qrs},\\[2pt]
  \label{eq:K-prop5}
  K_m{}^{[IJ} K_{n}{}^{KL]} K_p{}^{IJ} K_q{}^{KL} =&\, 16 \mathring
  g_{m(p} \mathring g_{nq)},\\[2pt]
  \label{eq:K-prop6}
  K_{mp}{}^{[IJ} K^{p\,KL]} K_{nq}{}^{IJ} K^{q\,KL} =&\, 192 \mathring
  g_{mn},\\[2pt]
  \label{eq:K-prop7}
  K_{[mn}{}^{[IJ} K_{p]}{}^{KL]} K^{[qr\,IJ} K^{s]KL} =&\, 32
  \delta_{mnp}^{qrs}.
\end{align}
whereas all other contractions, such as
\begin{equation}
  \label{eq:K-Gamma-0}
  \Gamma^m{}_{[AB} \Gamma^n{}_{CD]} \Gamma_{pq\,AB} \Gamma^q{}_{CD} =
  0,\qquad  K_m{}^{[IJ} K_{n}{}^{KL]} K_{pq}{}^{IJ} K^{q\,KL} = 0 
\end{equation}
vanish identically.

Finally, it is convenient to define the selfdual tensor
\begin{equation}
  \label{eq:K4}
  K^{IJKL} = K_m{}^{[IJ} K^{m\,KL]},
\end{equation}
which satisfies \cite{Godazgar:2015qia}
\begin{align}
  \label{eq:K-Identitites1}
  K^{IJKP} K_{LMNP} &= 6 \delta^{IJK}_{LMN} + 9 \delta^{[I}{}_{[L}
  K^{JK]}{}_{MN]},\\
  \label{eq:K-Identitites2}
  K^{[IJKL} K^{M]NPQ} &= \frac{1}{5} \epsilon^{IJKLMNPQ} + 12
  K^{[IJK}{}_{[N} \delta^{L}{}_{P} \delta^{M]}{}_{Q]},\\
  \label{eq:K-Identitites3}
  K^{m\,IJ} K^{n\,KL} K_{mn}{}^{MN} &= \, 8 \delta^{[I}{}_{[K}
  \delta^{J][M} \delta^{N]}{}_{L]} + 4 \delta^{[M}{}_{[I}
  K^{N]}{}_{J]KL} + 4\delta^{[K}{}_{[M} K^{L]}{}_{N]IJ} -
  4\delta^{[I}{}_{[K} K^{J]}{}_{L]MN}.
\end{align}

\bibliographystyle{utphys}
\bibliography{literature}

\end{document}